\documentclass[twocolumn]{article}
\usepackage{balance}
\usepackage{mathtools}
\usepackage{needspace}
\usepackage[ruled,vlined,linesnumbered]{algorithm2e}
\makeatletter
\gdef\@copyrightpermission{
  \begin{minipage}{0.2\columnwidth}
   \href{https://creativecommons.org/licenses/by/4.0/}{\includegraphics[width=0.90\textwidth]{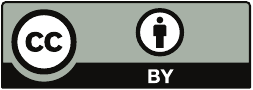}}
  \end{minipage}\hfill
  \begin{minipage}{0.8\columnwidth}
   \href{https://creativecommons.org/licenses/by/4.0/}{This work is licensed under a Creative Commons Attribution International 4.0 License.}
  \end{minipage}
  \vspace{1pt}
}
\makeatother

\usepackage{amsmath,amsfonts,amssymb,amsthm}
\usepackage{comment}
\usepackage{natbib}
\let\cite\citep
\usepackage{booktabs}
\usepackage{xcolor}
\usepackage[margin=0.75in, top=0.75in, bottom=1.0in]{geometry}
\usepackage{titlesec}

\titleformat{\paragraph}[runin]
  {\normalfont\bfseries} 
  {\theparagraph}        
  {1em}                  
  {}                     
  [{\bfseries.}]     
\usepackage{newtxtext, newtxmath} 

\usepackage[hidelinks]{hyperref}

\DeclareMathOperator*{\argmax}{arg\,max}
\DeclareMathOperator*{\argmin}{arg\,min}

\newtheorem{theorem}{Theorem}
\newtheorem{lemma}{Lemma}
\newtheorem{proposition}{Proposition}

\theoremstyle{definition}
\newtheorem{definition}{Definition}
\newtheorem{assumption}{Assumption}
\newtheorem{example}{Example}

\title{Stackelberg Games with Convex Risk Measures}

\author{
   Arunesh Sinha\\
   Management Science and Information Systems\\
   Rutgers University\\
   \texttt{arunesh.sinha@rutgers.edu}
 }
\date{}

\begin{document}

\maketitle

\begin{abstract}
Stackelberg games have important real-world applications in which reliable commitments require accounting for multiple sources of uncertainty.
We study finite Stackelberg games in which environmental uncertainty makes each mixed strategy profile induce a random reward, and each player evaluates that reward with a convex monetary risk criterion. In reward form, it is a concave, monotone, translation-equivariant utility and also represented as penalized expected reward over distributions of environmental outcomes. We provide two conditions under which the follower has a unique continuous mixed response, which leads to inducible optimistic equilibrium. We show that even approximate equilibrium computation is NP Hard. We then derive an exact mathematical program with equilibrium constraints for the problem.
We also develop an action entropy smoothed implicit-gradient method and prove an end-to-end additive approximation guarantee under the assumption that the gradient method reaches global optimality. We validate our scalability and other results with experiments on synthetic and real data.
\end{abstract}


\maketitle

\section{Introduction}

Research on Stackelberg games spans prominent algorithmic developments~\cite{conitzer2006computing,jain2010security,cerny2021computing} and real-world applications~\cite{fang2016deploying,pita2008effective,shieh2012protect}. In real-world applications, reliable commitments require accounting for uncertainty in payoff estimates, strategy execution, attacker observations, and behavioral models~\cite{yin2011riskaverse,kiekintveld2013interval,nguyen2014regret,nguyen2014unified}. Within Stackelberg models, particular forms of uncertainty have been addressed through distributionally robust commitments and robustness to approximately rational follower responses~\cite{Liu2018-ng,ananthanarayanan2022computingoptimaldistributionallyrobuststrategy,gan2025robust}. To our knowledge, how player-specific attitudes toward payoff uncertainty shape Stackelberg commitment remains underexplored. We fill this gap by allowing each player to evaluate the random payoff generated by the environment according to its own risk attitude.

Convex-risk utilities provide a unified model of risk-sensitive preferences that allows attitudes toward uncertainty to vary with the stakes. We adopt this risk model within the Stackelberg game framework. In our model, at the time strategies are chosen, the environmental condition remains unrealized; consequently, each strategy profile yields a random payoff for each player. \emph{Each player evaluates this payoff using its individual convex-risk utility}. Unlike standard Stackelberg games, the convex-risk utility results in the follower’s objective becoming nonlinear in its mixed strategy; consequently, the follower's best response may be mixed and may be strictly preferred to every pure strategy. We develop structural and computational results for this general class of Convex-risk Utilities based Stackelberg Games (CUSGs) and use entropic utility as the canonical example in this class.

As part of structural results, we analyze equilibrium in CUSGs. For equilibrium in Stackelberg games, two notions are important: (1) optimistic (resp., pessimistic) equilibrium assumes that, among multiple best responses, the follower selects one maximizing (resp., minimizing) the leader’s utility, and (2) inducibility asks whether a selected response can be approached by unique follower best responses induced by arbitrarily small perturbations of the leader’s commitment. As in standard Stackelberg games, an optimistic equilibrium exists in CUSGs, whereas a pessimistic one may not. However, \emph{inducibility fails for general CUSGs}. We identify two conditions under which every optimistic equilibrium is inducible and show that the resulting class includes CUSGs with entropic utility.

As part of our computational results, we show that even computing approximate equilibria in CUSGs is \emph{NP-hard}. Then, we present a mathematical program with equilibrium constraint (MPEC) \emph{formulation to compute equilibrium} for general CUSGs (under the sufficient conditions mentioned above), with an easier form when the follower employs entropic utility. We also \emph{propose an implicit gradient-based method} for equilibrium computation that builds upon entropy regularization to allow for differentiability, and prove an end-to-end approximate equilibrium guarantee under the assumption that the gradient method reaches global optimality.

Finally, we conduct experiments in a Stackelberg security game domain based on synthetic data as well as a wildlife protection domain based on real data from Lob\'ek\'e National Park~\cite{krever2025guard}. With synthetic data, under a one-hour limit per game instance, the gradient method and mathematical programming attain comparable leader utilities on small and medium instances, while neither method uniformly dominates on the largest instances. The real-data experiments reveal that more risk-averse attackers distribute attack probability across more targets, leading the defender to reallocate patrol coverage, and that optimizing for the players' risk preferences can improve leader utility relative to a risk-neutral strategy. Overall, our work provides a principled framework for risk considerations in Stackelberg games.

\section{Related Work}\label{sec:related-work}

\paragraph{Stackelberg Games under Uncertainty.}
Prior work addresses payoff and model uncertainty in Stackelberg security games through interval uncertainty and regret-based robust optimization~\cite{yin2011riskaverse,kiekintveld2013interval,nguyen2014regret,nguyen2014unified}. Distributionally robust commitment protects against ambiguity about the follower's utility model~\cite{ananthanarayanan2022computingoptimaldistributionallyrobuststrategy}, while \citet{mutzari2022robust} seek solutions that remain stable under perturbations of the attacker's utility parameters in multi-defender security games. Robust Stackelberg equilibrium instead protects the leader against approximately optimal follower responses~\cite{gan2025robust}. Related learning work in games considers noisy payoff observations, stochastic demand and production costs, and observable payoff-relevant contexts~\cite{bai2021sample,cesabianchi2023learning,harris2024regret,balcan2026nearly}. These prior works ask how the leader should commit when payoff parameters or their distribution are imperfectly known, estimated from noisy data, or when the follower is approximately optimal. Our uncertainty remains even when the game and some samples of environmental distribution are known: because the outcome is realized after the strategies are chosen, each strategy profile induces a random reward for each player. CUSGs allow each player to evaluate its random reward using an individual convex-risk utility. Unlike security game models of risk-averse attackers or an entropic-risk defender facing quantal-response attackers~\cite{qian2015robust,chicoisne2016risk,chicoisne2023algorithms}, our follower itself optimizes a convex-risk utility, making mixed responses and their inducibility central. Further away, recent work studies coherent-risk utilities in simultaneous-move games~\cite{gangwani2026data} and distributional robustness in continuous Stackelberg games~\cite{Liu2018-ng}; we instead study inducible equilibria in finite action Stackelberg games under convex risk utilities.

\paragraph{Computation of Stackelberg Equilibria.}
In finite normal-form games, optimal commitment can be computed by enumerating the follower's pure best responses and solving a linear program for each response~\cite{conitzer2006computing}. Security game algorithms exploit additional structure through mixed-integer formulations, compact coverage representations, and branch-and-price methods~\cite{paruchuri2008efficient,kiekintveld2009computing,jain2010security}. Quantal-response models instead prescribe a smooth follower response and use nonlinear optimization, transformations, or piecewise approximations~\cite{yang2012computing,cerny2021computing,mai2024tackling,li2023provable}. Recent methods for differentiable Stackelberg games compute gradients through follower equilibria using optimality conditions or implicit differentiation~\cite{fiez2020implicit,goktas2022robust,wang2022coordinating,loftin2024uncoupled}. CUSGs differ in that a mixed follower response can be a risk-utility maximizer and hence, pure-response enumeration does not characterize the leader's problem. 
We derive the exact MPEC induced by this convex-risk response and an entropy-smoothed implicit-gradient method for scalable computation.

\section{Preliminaries and Model}\label{sec:convex-preliminaries}

\paragraph{Convex-Risk Utilities}

Let $\mathcal{X}$ be the space of random variables defined on a finite probability space. Define convex-risk utility as
\begin{definition}[Convex-risk utility]\label{def:convex-monetary-utility}
\setlength{\topsep}{0pt} 
A functional $\rho:\mathcal{X}\to\mathbb{R}$ is a normalized convex-risk utility if, for all $X,Y\in\mathcal{X}$:
\begin{enumerate}
\item $\rho(\lambda X+(1-\lambda)Y)\geq\lambda\rho(X)+(1-\lambda)\rho(Y)$ for every $\lambda\in[0,1]$;
\item (monotonicity) $X\leq Y$ almost surely implies $\rho(X)\leq\rho(Y)$;
\item (translation equivariance) $\rho(X+c)=\rho(X)+c,$ $\forall c\in\mathbb{R}$; and
\item $\rho(0)=0$.
\end{enumerate}
\end{definition}
The reader may note that the first condition is a concavity condition under our reward convention; yet we call $\rho$ a convex-risk utility to keep the connection with the cost convention in risk literature~\cite{ruszczynski2006optimization}.
Write the support of the finite probability space as $\{\xi_1,\ldots,\xi_K\}$, where $\xi_k$ has probability $p_k>0$ and $\sum_{k=1}^K p_k=1$. Every $X\in\mathcal{X}$ can then be identified with its realization vector $\mathbf{z}=(X(\xi_k))_{k=1}^K$. 
Then, every $\rho$ is continuous and admits a penalized representation~\cite{foellmer2010convex,foellmer2011entropic}
\begin{equation}\label{eq:convex-penalty-representation}\begin{aligned}&\rho(X)=\min_{\mathbf{q}\in\Delta_K}\Big\{\sum_{k=1}^Kq_kX(\xi_k)+\alpha_\rho(\mathbf{q})\Big\},\\ &\mbox{where } \alpha_\rho(\mathbf{q})=\sup_{X\in\mathcal{X}}\Big\{\rho(X)-\sum_{k=1}^Kq_kX(\xi_k)\Big\},\end{aligned}\end{equation}
and $\alpha_\rho:\Delta_K\to[0,+\infty]$ is convex and lower semicontinuous. The probability vector $\mathbf{q}$ reweights the realizations $\xi_1,\ldots,\xi_K$, while $\alpha_\rho(\mathbf{q})$ assigns a penalty to such reweighting. Next, we provide a canonical convex risk utility that we use throughout this paper.
For a risk-sensitivity parameter $\theta>0$, the \emph{entropic utility}~\cite{foellmer2011entropic} of a random reward $X\in\mathcal{X}$ is
\begin{equation}\label{eq:entropic-utility}\rho_\theta(X)=-\frac{1}{\theta}\log\Big(\sum_{k=1}^Kp_k\exp\bigl(-\theta X(\xi_k)\bigr)\Big).\end{equation}
A larger $\theta$ gives greater weight to low-reward realizations. 
For all $X\in\mathcal{X}$, the penalized distribution form of the entropic utility is
\begin{equation}\label{eq:entropic-dual}\rho_\theta(X)=\min_{\mathbf{q}\in\Delta_K}\Big\{\sum_{k=1}^Kq_kX(\xi_k)+\frac{1}{\theta}\sum_{k=1}^Kq_k\log\frac{q_k}{p_k}\Big\},\end{equation}
with $0\log0=0$. The minimizer is unique and satisfies
\begin{equation}\label{eq:entropic-weights}q_k(X)=\frac{p_k\exp\bigl(-\theta X(\xi_k)\bigr)}{\sum_{k'=1}^Kp_{k'}\exp\bigl(-\theta X(\xi_{k'})\bigr)},\qquad k=1,\ldots,K.\end{equation}
Another convex-risk utility example is utility-based shortfall, which is shown in Appendix~\ref{app:additionalexample}.



\paragraph{Convex Utility Stackelberg Game (CUSG)}

Consider a leader $L$ and a follower $F$ with finite action sets $A_L$ and $A_F$ of sizes $n_L$ and $n_F$, respectively. Their mixed strategies are denoted by $\mathbf{x}\in\Delta(A_L)$ and $\mathbf{y}\in\Delta(A_F)$, respectively. Environmental uncertainty is represented by a random element $\xi$ whose distribution assigns probabilities $p_k>0$ to samples $\xi_1,\ldots,\xi_K$, with $\sum_{k=1}^Kp_k=1$. At pure action pair $(i,j)$ and realization $\xi_k$, player $\nu\in\{L,F\}$ receives the scalar reward $u_\nu(i,j\mid\xi_k)\in[0,1]$. 
A mixed profile induces the scalar-valued random reward $G_\nu(\mathbf{x},\mathbf{y}\mid\xi)$, whose realization at $\xi_k$ is
\begin{equation}\label{eq:convex-sample-payoff}g_{\nu,k}(\mathbf{x},\mathbf{y})=\sum_{i\in A_L}\sum_{j\in A_F}x_i y_j u_\nu(i,j\mid\xi_k),\end{equation}
and the vector $\mathbf{g}_\nu(\mathbf{x},\mathbf{y})=(g_{\nu,1}(\mathbf{x},\mathbf{y}),\ldots,g_{\nu,K}(\mathbf{x},\mathbf{y}))$ is the $K$-sample representation of that random variable. Every $g_{\nu,k}(\mathbf{x},\mathbf{y})$ lies in $[0,1]$. Player $\nu$ evaluates the scalar-valued random reward by
\begin{equation}\label{eq:convex-risk-adjusted-payoff}U_\nu(\mathbf{x},\mathbf{y})=\rho_\nu\!\left(G_\nu(\mathbf{x},\mathbf{y}\mid\xi)\right).\end{equation}
Thus the player's own mixed strategy randomization is averaged for environmental realization, while $\rho_\nu$ evaluates the remaining environmental uncertainty. 
Monotonicity, translation equivariance, and normalization give $\rho_\nu(0)=0$, $\rho_\nu(1)=1$, and therefore $U_\nu(\mathbf{x},\mathbf{y})\in[0,1]$.
The follower observes $\mathbf{x}$ and chooses from
$BR_F(\mathbf{x})=\argmax_{\mathbf{y}\in\Delta(A_F)}U_F(\mathbf{x},\mathbf{y}).$
As $\mathbf{g}_F(\mathbf{x},\cdot)$ is affine and the utility $U_F$ is concave, the follower problem is a convex optimization problem. 
We show in Example~\ref{ex:entropic-mixed-response} in Appendix~\ref{app:additionalexample} that a best response may be uniquely mixed and strictly preferred to every pure strategy. A mixed best response is economically meaningful: it hedges environmental payoff variation across follower actions.

\section{Equilibrium and Inducibility}


The optimistic and pessimistic Stackelberg values are, respectively,
\begin{equation*}\label{eq:optimistic-pessimistic-values}V_L^+=\max_{\mathbf{x}\in\Delta(A_L)}\max_{\mathbf{y}\in BR_F(\mathbf{x})}U_L(\mathbf{x},\mathbf{y}),\; V_L^-=\sup_{\mathbf{x}\in\Delta(A_L)}\min_{\mathbf{y}\in BR_F(\mathbf{x})}U_L(\mathbf{x},\mathbf{y}).\end{equation*}
Thus, when the follower has multiple optimal responses, the optimistic convention selects one maximizing the leader's utility, whereas the pessimistic convention selects one minimizing it. The supremum in the definition of $V_L^-$ is necessary because a maximizing leader commitment need not exist. The following result extends a similar known result in standard Stackelberg games to CUSGs.
\begin{proposition}[Existence of an optimistic equilibrium]\label{prop:general-cusg-existence}
Every CUSG admits an optimistic Stackelberg equilibrium. In contrast, a pessimistic Stackelberg equilibrium need not exist.
\end{proposition}
The existence of an optimistic equilibrium does not by itself justify favorable tie-breaking. The favorable follower response may fail to arise, even approximately, as a unique response to nearby leader commitments. Motivated by the notion of inducibility in Stackelberg games~\cite{vonstengelzamir2004} and the refined-best-response construction of \citet{balkenborg2013refined}, we use the sequence-based extension to continuous strategy spaces noted by \citet{gordon2026effective}.

\begin{definition}[Inducibility]\label{def:local-inducibility}
For $\mathbf{x}^*\in\Delta(A_L)$, a response $\mathbf{y}^*\in BR_F(\mathbf{x}^*)$ is \emph{inducible at $\mathbf{x}^*$} if either $BR_F(\mathbf{x}^*)=\{\mathbf{y}^*\}$ or there exist sequences $\mathbf{x}^{n}\in\Delta(A_L)\setminus\{\mathbf{x}^*\}$ and $\mathbf{y}^{n}\in\Delta(A_F)$ such that $\mathbf{x}^{n}\to\mathbf{x}^*$, $\mathbf{y}^{n}\to\mathbf{y}^*$, and $BR_F(\mathbf{x}^{n})=\{\mathbf{y}^{n}\}$ for every $n$. A CUSG has \emph{inducible optimistic equilibria} if every optimistic Stackelberg pair is inducible.
\end{definition}


The general CUSG does not have inducible equilibria. As an example for this, consider
Mean-semideviation (MSD)~\cite{ruszczynski2006optimization}, which is the convex-risk utility given as
\begin{equation}\label{eq:msd-inducibility-utility}
\rho_{\mathrm{MSD}}(X)=\mathbb{E}[X]-\gamma_s\mathbb{E}\!\left[\left(\mathbb{E}[X]-X\right)_+\right],\qquad \gamma_s\in[0,1].
\end{equation}

\begin{example}[A noninducible optimistic MSD response]\label{ex:msd-noninducible-response}
Let $A_L=\{L_1,L_2\}$ and $A_F=\{a,b\}$, with $\mathbf{x}=(p,1-p)$ and $\mathbf{y}=(q,1-q)$. There are two equally likely samples, and both players use MSD with $\gamma_s=1$. The follower payoff matrices are
\begin{equation*}\label{eq:msd-noninducible-follower-matrices}
\bigl(u_F(i,j\mid\xi_1)\bigr)_{ij}=\begin{pmatrix}7/8&1/4\\5/8&1/2\end{pmatrix},\quad\bigl(u_F(i,j\mid\xi_2)\bigr)_{ij}=\begin{pmatrix}3/8&1/4\\1/8&1/2\end{pmatrix}.
\end{equation*}
The resulting sample payoffs are $g_{F,1}(p,q)=1/2-p/4+q(1/8+p/2)$ and $g_{F,2}(p,q)=1/2-p/4+q(-3/8+p/2)$, so $g_{F,1}(p,q)-g_{F,2}(p,q)=q/2\geq0$. For a random reward with two equally likely realizations $z_1\geq z_2$, its mean is $(z_1+z_2)/2$ and its expected downside semideviation is $(z_1-z_2)/4$. Therefore, Equation~\eqref{eq:msd-inducibility-utility} with $\gamma_s=1$ gives $\rho_{\mathrm{MSD}}(z_1,z_2)=(z_1+3z_2)/4$. Therefore,
$
U_F(p,q)=\frac{3}{8}+\frac{1}{4}\left(p-\frac{1}{2}\right)(2q-1),
$
and hence $BR_F(p)=\{0\}$ for $p<1/2$, $BR_F(1/2)=[0,1]$, and $BR_F(p)=\{1\}$ for $p>1/2$.
Let the leader payoff matrices at $\xi_1$ and $\xi_2$ be $\left(\begin{smallmatrix}1&0\\1&0\end{smallmatrix}\right)$ and $\left(\begin{smallmatrix}0&1\\0&1\end{smallmatrix}\right)$, respectively. The leader's sample-payoff vector is $(q,1-q)$, so
$
U_L(q)=\frac{1}{2}-\frac{1}{2}\left|q-\frac{1}{2}\right|.
$
The unique optimistic Stackelberg pair is consequently $(p^*,q^*)=(1/2,1/2)$. It is not inducible because every unique follower response at a commitment different from $p^*$ belongs to $\{0,1\}$, so no sequence of such responses converges to $q^*=1/2$. 
\end{example}
The noninducibility in Example~\ref{ex:msd-noninducible-response} is caused by flatness of the follower's MSD utility. At $p=1/2$, pure action $a$ gives reward $3/4$ under $\xi_1$ and $1/4$ under $\xi_2$, whereas pure action $b$ gives reward $3/8$ under either realization. The difference between these two random rewards is therefore not almost surely constant. Nevertheless, we get $U_F(1/2,q)=3/8$ for every $q\in[0,1]$. Thus, $q=0$ and $q=1$ are distinct best responses that induce different random rewards, yet
$U_F(1/2,0)=U_F(1/2,1)=\max_{q\in[0,1]}U_F(1/2,q)=3/8$.


The following example shows that, even under entropic utility, which, as we show later, avoids the flatness of MSD, an optimistic follower response need not be inducible at the equilibrium strategy, motivating the second of the two conditions introduced afterward.

\begin{example}[A noninducible optimistic entropic response]\label{ex:entropic-noninducible-response}
Let $A_L=\{L_1,L_2\}$ and $A_F=\{a,b\}$, with $\mathbf{x}=(p,1-p)$ and $\mathbf{y}=(q,1-q)$. There are two equally likely samples, and rows and columns index leader and follower actions, respectively. In both samples, the follower payoff matrix is
\begin{equation*}\bigl(u_F(i,j\mid\xi_k)\bigr)_{ij}=\begin{pmatrix}1&0\\0&1\end{pmatrix},\qquad k=1,2.\end{equation*}
Thus the follower receives the same reward $(1-p)+(2p-1)q$ in both samples. Entropic utility therefore equals this deterministic reward, giving $BR_F(p)=\{0\}$ for $p<1/2$, $BR_F(1/2)=[0,1]$, and $BR_F(p)=\{1\}$ for $p>1/2$. Let the leader's payoff matrices at $\xi_1$ and $\xi_2$ be $\left(\begin{smallmatrix}1&0\\1&0\end{smallmatrix}\right)$ and $\left(\begin{smallmatrix}0&1\\0&1\end{smallmatrix}\right)$, respectively.
The leader's 2-sample-payoff realization vector is $(q,1-q)$. Its entropic utility $U_L(q)=-\theta_L^{-1}\log((e^{-\theta_Lq}+e^{-\theta_L(1-q)})/2)$ is symmetric and strictly concave in $q$, and is therefore uniquely maximized at $q=1/2$. Hence $(p^*,q^*)=(1/2,1/2)$ is the unique optimistic Stackelberg pair.
This pair is not inducible. Whenever the follower response is unique, it belongs to $\{0,1\}$, so no sequence of unique responses can converge to $q^*=1/2$. Consequently, the optimistic response $q^*=1/2$ is not inducible at $p^*=1/2$, although it belongs to $BR_F(1/2)=[0,1]$. 
\end{example}

The optimistic response is not inducible in the above example because, at $p=1/2$, $g_{F,1}(\mathbf{x},\mathbf{y})=g_{F,2}(\mathbf{x},\mathbf{y})=1/2$ for every $q\in[0,1]$. Hence $g_F(\mathbf{x},\mathbf{y}\mid\xi)=1/2$ and $U_F(\mathbf{x},\mathbf{y})=\rho_F(1/2)=1/2$, so even entropic utility cannot distinguish among the follower strategies. 

The above examples show that, more generally, multiple follower best responses mean that, for some $\mathbf{x}$, there exist $\mathbf{y}\neq\mathbf{y}'$ satisfying
\begin{equation}
U_F(\mathbf{x},\mathbf{y})=U_F(\mathbf{x},\mathbf{y}')=\max_{\widetilde{\mathbf{y}}\in\Delta(A_F)}U_F(\mathbf{x},\widetilde{\mathbf{y}}).
\end{equation}
This can occur for two reasons. First, the responses may induce different random rewards, $\mathbb{P}(G_F(\mathbf{x},\mathbf{y}\mid\xi)\neq G_F(\mathbf{x},\mathbf{y}'\mid\xi))>0$, that nevertheless have equal utility. Second, they may induce the same random reward, $G_F(\mathbf{x},\mathbf{y}\mid\xi)=G_F(\mathbf{x},\mathbf{y}'\mid\xi)$ almost surely, in which case no utility functional applied to that reward can distinguish them. These two sources motivate the conditions \emph{strict concavity modulo translations} and \emph{active-tie independence}, respectively.
Strict concavity modulo translations addresses the first source.
\begin{definition}[Strict concavity modulo translations]\label{def:strict-concavity-modulo-translations}
A convex-risk utility $\rho$ is strictly concave modulo translations if, whenever $X-Y$ is not almost surely constant and $\lambda\in(0,1)$,
\begin{equation}\label{eq:strict-concavity-modulo-translations}\rho(\lambda X+(1-\lambda)Y)>\lambda\rho(X)+(1-\lambda)\rho(Y).\end{equation}
\end{definition}
To see the implication of strict concavity modulo translations, let $\mathbf{y},\mathbf{y}'\in BR_F(\mathbf{x})$ and write $X=G_F(\mathbf{x},\mathbf{y}\mid\xi)$ and $Y=G_F(\mathbf{x},\mathbf{y}'\mid\xi)$. Affinity in the follower strategy gives $G_F(\mathbf{x},(\mathbf{y}+\mathbf{y}')/2\mid\xi)=(X+Y)/2$. If $X-Y$ is not almost surely constant, Definition~\ref{def:strict-concavity-modulo-translations} gives
\begin{equation}
U_F\!\left(\mathbf{x},(\mathbf{y}+\mathbf{y}')/2\right)>(1/2)U_F(\mathbf{x},\mathbf{y})+(1/2)U_F(\mathbf{x},\mathbf{y}'),
\end{equation}
contradicting the optimality of $\mathbf{y}$ and $\mathbf{y}'$. Hence $X-Y=c$ almost surely for some $c$. Translation equivariance gives $U_F(\mathbf{x},\mathbf{y})=U_F(\mathbf{x},\mathbf{y}')+c$; because both strategies are best responses, their utilities are equal, so $c=0$. Therefore, $X=Y$ almost surely. 

Next, since $p_k>0$ for every $k$, $X=Y$ is equivalent to $\mathbf{g}_F(\mathbf{x},\mathbf{y})=\mathbf{g}_F(\mathbf{x},\mathbf{y}')$ (recall, $\mathbf{g}_F$ is a vector of size $K$). However, this does not rule out $\mathbf{y} \neq \mathbf{y}'$. Affinity in the follower strategy gives
\begin{equation}
\sum_{j\in A_F}(y_j-y_j')\mathbf{g}_F(\mathbf{x},\mathbf{e}_j)=\mathbf{0},\qquad \sum_{j\in A_F}(y_j-y_j')=0.
\end{equation}
Thus, under strict concavity modulo translations, two distinct strategies can still both be follower best responses only if the sample-payoff vectors $\mathbf{g}_F(\mathbf{x},\mathbf{e}_j)$ induced by the pure follower actions supporting them are affinely dependent. The next condition rules out this remaining source of multiplicity, noting that requiring affine independence of $\{\mathbf{g}_F(\mathbf{x},\mathbf{e}_j):j\in A_F\}$ would be unnecessarily strong.

We want to rule out two distinct responses $\mathbf{y},\mathbf{y}'\in BR_F(\mathbf{x})$ satisfying $\mathbf{g}_F(\mathbf{x},\mathbf{y})=\mathbf{g}_F(\mathbf{x},\mathbf{y}')$.
The penalized-distribution representation of $\rho_F$ in Equation~\eqref{eq:convex-penalty-representation}, with penalty function $\alpha_F$, identifies which pure actions can occur in an optimal follower response. For a commitment $\mathbf{x}$ and a distribution $\mathbf{q}\in\Delta_K$ satisfying $\alpha_F(\mathbf{q})<+\infty$, define 
\begin{equation}\label{eq:active-follower-actions}J_F(\mathbf{x},\mathbf{q})=\argmax_{j\in A_F}\sum_{k=1}^Kq_k g_{F,k}(\mathbf{x},\mathbf{e}_j).\end{equation}
Thus, $J_F(\mathbf{x},\mathbf{q})$ contains the pure follower actions tied for the largest $\mathbf{q}$-weighted payoff. 
{\color{black}For fixed $\mathbf{x}$, the follower's dual problem is
\begin{equation}\label{eq:follower-minimax-dual}\min_{\mathbf{q}\in\Delta_K}\left\{\alpha_F(\mathbf{q})+\max_{j\in A_F}\sum_{k=1}^Kq_k g_{F,k}(\mathbf{x},\mathbf{e}_j)\right\}.\end{equation}
Finite-dimensional minimax duality equates its optimal value with $\max_{\mathbf{y}\in\Delta(A_F)}U_F(\mathbf{x},\mathbf{y})$. Moreover, for any minimizer $\mathbf{q}_F$ of Problem~\eqref{eq:follower-minimax-dual}, every $\mathbf{y}\in BR_F(\mathbf{x})$ satisfies $\operatorname{supp}(\mathbf{y})\subseteq J_F(\mathbf{x},\mathbf{q}_F)$, as established in the proof of Theorem~\ref{thm:convex-risk-attainment}. Thus, affine independence on this one common tied set suffices to distinguish the strategies of any two best responses that generate identical sample rewards. This motivates the following condition.\par}


\begin{assumption}[Active-tie independence]\label{ass:active-tie-independence}
{\color{black}For every $\mathbf{x}\in\Delta(A_L)$, there exists a minimizer $\mathbf{q}_F$ of Problem~\eqref{eq:follower-minimax-dual} such that the vectors $\{\mathbf{g}_F(\mathbf{x},\mathbf{e}_j):j\in J_F(\mathbf{x},\mathbf{q}_F)\}$ are affinely independent. Explicitly, if scalars $c_j$, $j\in J_F(\mathbf{x},\mathbf{q}_F)$, satisfy $\sum_jc_j=0$ and $\sum_jc_j\mathbf{g}_F(\mathbf{x},\mathbf{e}_j)=\mathbf{0}$, then $c_j=0$ for every $j\in J_F(\mathbf{x},\mathbf{q}_F)$.\par}
\end{assumption}

{\color{black}The selected distribution $\mathbf{q}_F$ may depend on $\mathbf{x}$. Independence is required under only one dual-optimal distribution at each leader strategy. In particular, the condition bounds $|J_F(\mathbf{x},\mathbf{q}_F)|$, not the tied sets under arbitrary distributions. In Example~\ref{ex:entropic-noninducible-response}, at $p=1/2$ the dual optimizer is the nominal distribution, both pure actions belong to its tied set, and their payoff vectors coincide; hence the condition fails. By contrast, the two pure-action payoff vectors in Example~\ref{ex:msd-noninducible-response} and Example~\ref{ex:cvar-noninducible-response} are distinct for every leader strategy, so every tied subset is affinely independent and the condition holds despite the failure of strict concavity modulo translations.\par}

The next result establishes the equilibrium attainment and inducibility under the conditions above. The proof relies on the arguments presented above for unique follower response coupled with standard real analysis techniques.

\begin{theorem}[Unique response and attained commitment]\label{thm:convex-risk-attainment}
Suppose $\rho_L$ and $\rho_F$ are convex-risk utilities, $\rho_F$ is strictly concave modulo translations, and Assumption~\ref{ass:active-tie-independence} holds. Then $BR_F(\mathbf{x})$ is a singleton for every $\mathbf{x}\in\Delta(A_L)$. Its unique element $\mathbf{y}^*(\mathbf{x})$ satisfies $|\operatorname{supp}(\mathbf{y}^*(\mathbf{x}))|\leq\min\{n_F,K\}$, and the map $\mathbf{y}^*:\Delta(A_L)\to\Delta(A_F)$ is continuous. Consequently,
\begin{equation}\label{eq:convex-risk-stackelberg-value}V=\max_{\mathbf{x}\in\Delta(A_L)}U_L\!\left(\mathbf{x},\mathbf{y}^*(\mathbf{x})\right)\end{equation}
is attained, and the optimistic and pessimistic Stackelberg values coincide with $V$. Thus, the follower response at every Stackelberg equilibrium is inducible.
\end{theorem}


For entropic utility, Lemma~\ref{lem:entropic-properties} below implies that Theorem~\ref{thm:convex-risk-attainment} applies whenever Assumption~\ref{ass:active-tie-independence} holds, yielding inducible equilibria. 
\begin{lemma}
\label{lem:entropic-properties}
Entropic utility is strictly concave modulo translations. 
\end{lemma}
\paragraph{Other Risk Measures} As a final note in structural results, CUSGs with coherent-risk utilities~\cite{ruszczynski2006optimization} need not be inducible; coherent risk measures are a subset of convex risk measures that additionally satisfy positive homogeneity: $\rho(tX)=t\rho(X)$ for $t>0$ and $X\in\mathcal{X}$. Popular risk measures such as Conditional Value at Risk (CVaR) and Mean-semideviation (MSD) are coherent. We show in Appendix~\ref{app:additionalexample} that positive homogeneity precludes strict concavity modulo translations. Example~\ref{ex:msd-noninducible-response} shows our claimed failure of inducibility for MSD, and Example~\ref{ex:cvar-noninducible-response} in the appendix shows the same for CVaR.

\section{Hardness of Approximation}\label{app:constant-additive-hardness}

This section shows approximation hardness in CUSGs. An $\epsilon$- Stackelberg equilibrium is a pair $(\widehat{\mathbf{x}},\widehat{\mathbf{y}})$ such that $\widehat{\mathbf{y}}=\mathbf{y}^*(\widehat{\mathbf{x}})$ is the exact follower response and $U_L(\widehat{\mathbf{x}},\widehat{\mathbf{y}})\geq V-\epsilon$, where $V$ is the optimal value in Equation~\eqref{eq:convex-risk-stackelberg-value}. We start by introducing the source problem.
\paragraph{Source Problem: Cubic Maximum Cut}

Let $G=(V,E)$ be an undirected $3$-regular graph with $V=\{1,\ldots,n\}$. A cut is a subset $C\subseteq V$, and its size is the number of edges having exactly one endpoint in $C$. Write $m=|E|=3n/2$ and
\begin{equation}\label{eq:cubic-max-cut-value}\operatorname{MC}(G)=\max_{C\subseteq V}\left|\bigl\{\{i,j\}\in E:i\in C,\ j\notin C\bigr\}\right|.\end{equation}
Berman and Karpinski~\cite{berman2001efficient} prove a constant inapproximability result for this problem, improving on their prior result~\cite{berman1999tighter}. In the form needed here, there are constants $r_{\mathrm{yes}}>r_{\mathrm{no}}$ such that it is NP-hard to distinguish graphs satisfying $\operatorname{MC}(G)/m\geq r_{\mathrm{yes}}$ from ones satisfying $\operatorname{MC}(G)/m\leq r_{\mathrm{no}}$. Denote this fixed separation by $\gamma_{\mathrm{cut}}=r_{\mathrm{yes}}-r_{\mathrm{no}}>0$; $\gamma_{\mathrm{cut}}$ can be taken to be $1/156 - 2\delta$ for any $\delta > 0$.

Towards our requirement, we first express $\operatorname{MC}(G)$ as a continuous optimization problem that will later be encoded by the follower response. For a binary vector $\mathbf{t}'\in\{0,1\}^n$ representing a cut, $|t'_i-t'_j|$ equals one exactly when edge $\{i,j\}$ has endpoints on opposite sides of the cut, and equals zero otherwise. Thus, $\sum_{\{i,j\}\in E}|t'_i-t'_j|$ is exactly the cut size. The same expression is well defined for every nonnegative vector $\mathbf t$ and provides a continuous extension of the cut objective. For a vector $\mathbf{t}\in\mathbb{R}_+^n$, define
\begin{equation}\label{eq:continuous-cut-and-overflow}S_G(\mathbf{t})=\sum_{\{i,j\}\in E}|t_i-t_j|,\qquad P(\mathbf{t})=\sum_{i=1}^n(t_i-1)_+.\end{equation}
The vector $\mathbf t$ may have coordinates outside the binary range $[0,1]$, so $P(\mathbf{t})$ defined above measures the total excess above one: coordinate $i$ contributes $t_i-1$ when $t_i>1$ and zero otherwise. 
The following lemma shows that maximum cut admits an exact continuous optimization formulation. The core idea is that the penalty $4P(\mathbf t)=4\sum_i(t_i-1)_+$ removes any benefit from coordinates above $1$; for $\mathbf t\in[0,1]^n$, $S_G(\mathbf t)$ equals the expected size of a threshold cut and therefore cannot exceed $\operatorname{MC}(G)$, while a maximum-cut indicator in $\{0,1\}^n$ attains equality in the lemma below.
\begin{lemma}[Continuous cubic maximum-cut identity]\label{lem:continuous-cubic-max-cut}
For every $3$-regular graph $G$ on $n$ vertices,
\begin{equation}\label{eq:continuous-cubic-max-cut-identity}\max_{\mathbf{t}\geq\mathbf{0},\ \mathbf{1}^{\top}\mathbf{t}\leq n}\left\{S_G(\mathbf{t})-4P(\mathbf{t})\right\}=\operatorname{MC}(G).\end{equation}
\end{lemma}

\paragraph{Reduction to an Entropic Stackelberg Game}

Fix a cubic maximum cut graph instance $G=(V,E)$. We construct a Stackelberg game with entropic utility. The leader has one action for every vertex and one additional dummy action:
$A_L=\{0\}\cup V.$
For any $\mathbf{x}\in\Delta(A_L)$, the coordinate $x_i$ is the probability assigned to vertex action $i\in V$, while $x_0$ is the probability assigned to the dummy action. Define $t_i=nx_i$ for each $i\in V$. Since $\sum_{i\in V}x_i=1-x_0\leq1$, the induced vector satisfies $\mathbf{t}\geq\mathbf{0}$ and $\mathbf{1}^{\top}\mathbf{t}\leq n$. Conversely, every nonnegative $\mathbf{t}$ satisfying this inequality is represented by $x_i=t_i/n$ and $x_0=1-\mathbf{1}^{\top}\mathbf{t}/n$. Thus, the leader simplex represents exactly the feasible region in Lemma~\ref{lem:continuous-cubic-max-cut}.
Set the fixed rational constants
\begin{equation}\label{eq:additive-hardness-fixed-constants}a=1/4,\qquad \lambda=1/16.\end{equation}
The follower actions are divided into three groups. For every undirected edge $\{i,j\}\in E$, include two ordered edge actions $e_{ij}$ and $e_{ji}$. For every vertex $i\in V$, include one node action $p_i$. Finally, include $M$ anchor actions $z_1,\ldots,z_M$, where the positive integer $M$ will be fixed later before the main theorem. Hence
\begin{equation}\label{eq:additive-hardness-follower-actions}A_F=\{e_{ij},e_{ji}:\{i,j\}\in E\}\cup\{p_i:i\in V\}\cup\{z_1,\ldots,z_M\}.\end{equation}
There is one environmental sample $\xi_s$ for every follower action $s\in A_F$, and the empirical distribution is uniform over these $K=2m+n+M$ samples. A follower strategy is $\mathbf{y}\in\Delta(A_F)$, and $y_s$ denotes the probability assigned to follower action $s$.

For leader action $l\in A_L$ and an edge sample $e_{ij}$, define the rational coefficient
\begin{equation}\label{eq:additive-hardness-cut-coefficients}D_{l,e_{ij}}=(a\lambda n/m)\left(\mathbf{1}_{\{l=i\}}-\mathbf{1}_{\{l=j\}}\right).\end{equation}
For a node sample $p_i$ and an anchor sample $z_r$, respectively, set
\begin{equation}\label{eq:additive-hardness-other-coefficients}D_{l,p_i}=(4a\lambda/m)\left(n\mathbf{1}_{\{l=i\}}-1\right),\qquad D_{l,z_r}=0.\end{equation}
Here $\mathbf{1}_{\{l=i\}}$ is the indicator of the event when the leader action is vertex $i$ and zero otherwise. The dummy action $l=0$ is distinct from every vertex, so every vertex indicator in Equation~\eqref{eq:additive-hardness-cut-coefficients} and Equation~\eqref{eq:additive-hardness-other-coefficients} is zero when $l=0$. For each sample $s\in A_F$, let $d_s(\mathbf{x})=\sum_{l\in A_L}x_l D_{l s}$ denote the coefficient $D_{l s}$ averaged over the leader commitment $\mathbf{x}$. Substituting $t_i=nx_i$ and $\sum_{l\in A_L}x_l=1$ gives
\begin{equation}\label{eq:additive-hardness-feature-scores}d_{e_{ij}}(\mathbf{x})=\frac{a\lambda}{m}(t_i-t_j),\quad d_{p_i}(\mathbf{x})=\frac{4a\lambda}{m}(t_i-1),\quad d_{z_r}(\mathbf{x})=0.\end{equation}
Define the follower's pure payoff at leader action $l$, follower action $r$, and sample $\xi_s$ as
\begin{equation}\label{eq:additive-hardness-follower-payoffs}u_F(l,r\mid\xi_s)= 1/2 +a\mathbf{1}_{\{r=s\}}-D_{l s}.\end{equation}
These payoffs satisfy assumed normalization. Indeed, $m=3n/2$ implies $|D_{l,e_{ij}}|\leq a\lambda n/m=1/96$. For an vertex sample, $D_{i,p_i}=4a\lambda(n-1)/m<1/24$, while $D_{l,p_i}=-4a\lambda/m$ for $l\neq i$ and its magnitude is at most $1/96$ for every 3-regular graph with $n\geq4$. So every payoff in Equation~\eqref{eq:additive-hardness-follower-payoffs} lies between $11/24$ and $73/96$, in particular in $[0,1]$.

For a mixed profile $(\mathbf{x},\mathbf{y})$, substituting Equation~\eqref{eq:additive-hardness-follower-payoffs} into the sample-payoff definition Equation~\eqref{eq:convex-sample-payoff} yields
\begin{equation}\label{eq:additive-hardness-follower-sample-payoff}g_{F,s}(\mathbf{x},\mathbf{y})=1/2+ay_s-d_s(\mathbf{x}),\qquad s\in A_F.\end{equation}
\emph{Both players use entropic utility} with fixed parameters $\theta_L=\theta_F=1$. For fixed $\mathbf{x}$, maximizing the follower utility in Equation~\eqref{eq:additive-hardness-follower-sample-payoff} is equivalent to minimizing
\begin{equation}\label{eq:additive-hardness-follower-exponential-program}\Phi_{\mathbf{x}}(\mathbf{y})=\sum_{s\in A_F}\exp\bigl(d_s(\mathbf{x})-ay_s\bigr)\qquad\text{over }\mathbf{y}\in\Delta(A_F).\end{equation}
The omitted factors $K^{-1}\exp(-1/2)$ are positive and independent of $\mathbf{y}$. Since $\Phi_{\mathbf{x}}$ is strictly convex in $\mathbf{y}$, its KKT conditions are necessary and sufficient and determine a unique follower response.
The following proposition characterizes the follower response and quantifies the effect of the anchor actions.

\begin{proposition}[Follower Threshold and Anchor Bound]\label{prop:additive-hardness-threshold-response}
For all leader strategy $\mathbf{x}$, the follower problem in Equation~\eqref{eq:additive-hardness-follower-exponential-program} has a unique solution $\mathbf{y}^*(\mathbf{x})$. There is a unique threshold $\tau(\mathbf{x})$ such that
\begin{equation}\label{eq:additive-hardness-threshold-response}y_s^*(\mathbf{x})=\frac{(d_s(\mathbf{x})-\tau(\mathbf{x}))_+}{a},\qquad \sum_{s\in A_F}(d_s(\mathbf{x})-\tau(\mathbf{x}))_+=a.\end{equation}
Define the zero-threshold version $\bar y_s(\mathbf{x})=(d_s(\mathbf{x})/a)_+$ and let $e(\mathbf{x})=-\tau(\mathbf{x})/a$. Then $\tau(\mathbf{x})<0$, every anchor action receives probability $y_{z_r}^*(\mathbf{x})=e(\mathbf{x})$, and
$0<e(\mathbf{x})\leq(1/M).$
Moreover, for every $s\in A_F$,
\begin{equation}\label{eq:additive-hardness-feature-error}y_s^*(\mathbf{x})=\Big(\frac{d_s(\mathbf{x})}{a}+e(\mathbf{x})\Big)_+,\quad 0\leq y_s^*(\mathbf{x})-\bar y_s(\mathbf{x})\leq e(\mathbf{x})\leq 1/M.\end{equation}
\end{proposition}

We complete the construction by defining the leader payoffs and relating the resulting leader value to the continuous maximum-cut objective. For every $l\in A_L$, $r\in A_F$, and sample $\xi_s$, set
\begin{equation}\label{eq:additive-hardness-leader-payoffs}u_L(l,r\mid\xi_s)=\begin{cases}1,&r\in\{e_{ij},e_{ji}\}\text{ for some }\{i,j\}\in E,\\0,&r=p_i\text{ for some }i\in V,\\1/2,&r\in\{z_1,\ldots,z_M\}.\end{cases}\end{equation}

\begin{proposition}[Maximum-Cut Value Encoding]\label{prop:additive-hardness-value-encoding}

For the follower response $\mathbf{y}^*(\mathbf{x})$, define
$Y_E(\mathbf{x})=\sum_{\{i,j\}\in E}\big(y_{e_{ij}}^*(\mathbf{x})+y_{e_{ji}}^*(\mathbf{x})\big)$, $\qquad Y_P(\mathbf{x})=\sum_{i=1}^ny_{p_i}^*(\mathbf{x}),\; Y_0(\mathbf{x})=\sum_{r=1}^My_{z_r}^*(\mathbf{x}).$
The leader receives the same reward in every sample, and its utility $U_L(\mathbf{x},\mathbf{y}^*(\mathbf{x}))$ is
\begin{equation}\label{eq:additive-hardness-leader-deterministic-value}W(\mathbf{x})=Y_E(\mathbf{x})+Y_0(\mathbf{x})/2 = 1/2+(Y_E(\mathbf{x})-Y_P(\mathbf{x}))/2.\end{equation}
Define $\bar Y_E(\mathbf{x})$ and $\bar Y_P(\mathbf{x})$ by replacing $y_s^*(\mathbf{x})$ with $\bar y_s(\mathbf{x})$ in the definitions of $Y_E(\mathbf{x})$ and $Y_P(\mathbf{x})$, and define the comparison value $\bar W(\mathbf{x})=1/2+(\bar Y_E(\mathbf{x})-\bar Y_P(\mathbf{x}))/2$. Then
$\bar Y_E(\mathbf{x})=\frac{\lambda}{m}S_G(\mathbf{t})$, $ \bar Y_P(\mathbf{x})=\frac{4\lambda}{m}P(\mathbf{t})$, $\bar W(\mathbf{x})=\frac{1}{2}+\frac{\lambda}{2m}\left(S_G(\mathbf{t})-4P(\mathbf{t})\right)$, 
which gives
\begin{equation}\label{eq:additive-hardness-intended-optimum}\max_{\mathbf{x}\in\Delta(A_L)}\bar W(\mathbf{x})=1/2 + (\lambda/{(2m)})\operatorname{MC}(G).\end{equation}
Moreover, uniformly over all $\mathbf{x}\in\Delta(A_L)$,
\begin{equation}\label{eq:additive-hardness-leader-approximation}|W(\mathbf{x})-\bar W(\mathbf{x})|\leq(2m+n)/(2M).\end{equation}
\end{proposition}

\paragraph{Structural Assumptions and Additive Separation}
Recall that we restrict our class of games to those that satisfy the conditions in Theorem~\ref{thm:convex-risk-attainment}, hence we verify the following as a sanity check
\begin{proposition}\label{prop:sanitycheck}
    The entropic utility CUSG constructed above satisfies Assumption~\ref{ass:active-tie-independence} and strict concavity modulo translations.
\end{proposition}
Next, choose
\begin{equation}\label{eq:additive-hardness-anchor-count}M\geq 8(2m+n)/(\lambda\gamma_{\mathrm{cut}}).\end{equation}
Because $G$ is a 3-regular graph, $m=3n/2$, we have $2m+n=4n$. Since $\lambda$ and $\gamma_{\mathrm{cut}}$ are fixed positive constants, the number of anchors is linear in the source instance size. All payoffs are rational with polynomial encoding length, and the full Stackelberg instance has polynomial size. Equations~\eqref{eq:additive-hardness-leader-approximation} and \eqref{eq:additive-hardness-anchor-count} give
\begin{equation}\label{eq:additive-hardness-uniform-error}\left|W(\mathbf{x})-\bar W(\mathbf{x})\right|\leq \lambda\gamma_{\mathrm{cut}}/{16}\qquad\text{for every }\mathbf{x}\in\Delta(A_L).\end{equation}

\begin{theorem}[Constant-additive hardness]\label{thm:constant-additive-hardness}
There exists a fixed constant $\epsilon_0>0$ such that, for every fixed $\epsilon\in(0,\epsilon_0]$, computing an $\epsilon$-Stackelberg equilibrium is NP-hard for CUSGs for which strict concave modulo translations and Assumption~\ref{ass:active-tie-independence} holds.
\end{theorem}

\begin{proof}
Given a graph instance $G$ for the cubic maximum-cut problem, construct the Stackelberg game instance as specified before the theorem, taking $M=\lceil 8(2m+n)/(\lambda\gamma_{\mathrm{cut}})\rceil$. The size argument preceding the theorem shows that this transformation is computable in polynomial time. Let $V(G)=\max_{\mathbf{x}}W(\mathbf{x})$ be the Stackelberg value of the constructed instance. If $G$ is a yes-instance of the maximum-cut problem, Equations~\eqref{eq:additive-hardness-intended-optimum} and \eqref{eq:additive-hardness-uniform-error} give
\begin{equation}\label{eq:additive-hardness-yes-value}V(G)\geq1/2+\lambda r_{\mathrm{yes}}/2-\lambda\gamma_{\mathrm{cut}}/16.\end{equation}
If $G$ is a no-instance, the same equations give
$V(G)\leq1/2+\lambda r_{\mathrm{no}}/2 + \lambda\gamma_{\mathrm{cut}}/16.$
The lower bound exceeds the upper bound  by
\begin{equation}\label{eq:additive-hardness-stackelberg-separation}(\lambda/2)(r_{\mathrm{yes}}-r_{\mathrm{no}})-\lambda\gamma_{\mathrm{cut}}/8 = 3\lambda\gamma_{\mathrm{cut}}/{8}.\end{equation}
The above uses  $\gamma_{\mathrm{cut}}=r_{\mathrm{yes}}-r_{\mathrm{no}}$. Set $\epsilon_0=\lambda\gamma_{\mathrm{cut}}/16$, which is a fixed positive constant independent of the graph size. Fix any $\epsilon\in(0,\epsilon_0]$.

Suppose a poly-time algorithm returns an $\epsilon$ Stackelberg equilibrium $(\widehat{\mathbf{x}},\widehat{\mathbf{y}})$. By definition, $\widehat{\mathbf{y}}=\mathbf{y}^*(\widehat{\mathbf{x}})$ and $W(\widehat{\mathbf{x}})=U_L(\widehat{\mathbf{x}},\widehat{\mathbf{y}})\geq V(G)-\epsilon$. The exact response $\mathbf{y}^*(\widehat{\mathbf{x}})$, and hence the leader value $W(\widehat{\mathbf{x}})=U_L(\widehat{\mathbf{x}},\mathbf{y}^*(\widehat{\mathbf{x}}))$, can be evaluated in polynomial time from the returned commitment. To compute $\mathbf{y}^*(\widehat{\mathbf{x}})$, recall that $a=1/4$ and that Equation~\eqref{eq:additive-hardness-threshold-response} requires $\sum_s(d_s(\widehat{\mathbf{x}})-\tau(\widehat{\mathbf{x}}))_+=a$. Sort the scores as $d_{(1)}(\widehat{\mathbf{x}})\geq\cdots\geq d_{(K)}(\widehat{\mathbf{x}})$. If precisely the first $k$ coordinates are strictly positive in $\mathbf{y}^*(\widehat{\mathbf{x}})$, then the simplex constraint requires
\begin{equation}\label{eq:additive-hardness-computed-threshold}
1=\sum_{j=1}^k\frac{d_{(j)}(\widehat{\mathbf{x}})-\tau_k}{a}
\quad\Longrightarrow\quad
\tau_k=\frac{\sum_{j=1}^k d_{(j)}(\widehat{\mathbf{x}})-a}{k}.
\end{equation}

The unique $k$ satisfying $d_{(k)}(\widehat{\mathbf{x}})>\tau_k\geq d_{(k+1)}(\widehat{\mathbf{x}})$, with the second inequality omitted when $k=K$, identifies the strictly positive coordinates and gives $\tau(\widehat{\mathbf{x}})=\tau_k$. Thus, after sorting the $K=|A_F|$ scores, compute $\tau_k$ from Equation~\eqref{eq:additive-hardness-computed-threshold} for every $k=1,\ldots,K$ and select the unique $k$ satisfying $d_{(k)}(\widehat{\mathbf{x}})>\tau_k\geq d_{(k+1)}(\widehat{\mathbf{x}})$. This gives $\tau(\widehat{\mathbf{x}})=\tau_k$ and computes $\mathbf{y}^*(\widehat{\mathbf{x}})$ in polynomial time, after which Equation~\eqref{eq:additive-hardness-leader-deterministic-value} gives $W(\widehat{\mathbf{x}})$.

In a yes-instance, Equation~\eqref{eq:additive-hardness-yes-value}, $\epsilon\leq\epsilon_0$, and $\epsilon_0=\lambda\gamma_{\mathrm{cut}}/16$ imply
\begin{equation}\label{eq:additive-hardness-yes-commitment}W(\widehat{\mathbf{x}})\geq V(G)-\epsilon\geq V(G)-\epsilon_0\geq\frac{1}{2}+\frac{\lambda}{4}(r_{\mathrm{yes}}+r_{\mathrm{no}})+\frac{\lambda\gamma_{\mathrm{cut}}}{8}.\end{equation}
The equality underlying the last expression is $\lambda r_{\mathrm{yes}}/2-\lambda\gamma_{\mathrm{cut}}/8=\lambda(r_{\mathrm{yes}}+r_{\mathrm{no}})/4+\lambda\gamma_{\mathrm{cut}}/8$. In a no-instance, $W$ satisfies
\begin{equation}\label{eq:additive-hardness-no-commitment}W(\widehat{\mathbf{x}})\leq V(G)\leq\frac{1}{2}+\frac{\lambda r_{\mathrm{no}}}{2}+\frac{\lambda\gamma_{\mathrm{cut}}}{16}=\frac{1}{2}+\frac{\lambda}{4}(r_{\mathrm{yes}}+r_{\mathrm{no}})-\frac{3\lambda\gamma_{\mathrm{cut}}}{16}.\end{equation}
The equality in the above follows from $\gamma_{\mathrm{cut}}=r_{\mathrm{yes}}-r_{\mathrm{no}}$. Equation~\eqref{eq:additive-hardness-yes-commitment} places $W(\widehat{\mathbf{x}})$ above the midpoint $1/2+\lambda(r_{\mathrm{yes}}+r_{\mathrm{no}})/4$ by at least $\lambda\gamma_{\mathrm{cut}}/8$, whereas Equation~\eqref{eq:additive-hardness-no-commitment} places it below that midpoint by at least $3\lambda\gamma_{\mathrm{cut}}/16$. Thus, comparing the computed value with this rational midpoint decides the cubic maximum-cut problem in poly time. Since $\epsilon\in(0,\epsilon_0]$ was arbitrary, computing an $\epsilon$-Stackelberg equilibrium is NP-hard for every fixed $\epsilon$ in this interval.
\end{proof}

\section{Equilibrium Computation Approaches}\label{sec:convex-exact-computation}

\paragraph{Mathematical programming based approach}

Under the optimistic equilibrium convention, the convex-risk Stackelberg problem is
\begin{equation}\label{eq:convex-risk-bilevel}\max_{\mathbf{x}\in\Delta(A_L),\,\mathbf{y}\in\Delta(A_F)}U_L(\mathbf{x},\mathbf{y})\quad\mathrm{s.t.}\quad \mathbf{y}\in\argmax_{\widetilde{\mathbf{y}}\in\Delta(A_F)}U_F(\mathbf{x},\widetilde{\mathbf{y}}).\end{equation}
Let $\alpha_F$ be the penalty representing $\rho_F$ in \eqref{eq:convex-penalty-representation}. For a fixed strategy profile, a minimizing distribution $\mathbf{q}_F$ satisfies
\begin{equation}\label{eq:follower-convex-risk-dual}U_F(\mathbf{x},\mathbf{y})=\min_{\mathbf{q}\in\Delta_K}\left\{\sum_{k=1}^Kq_kg_{F,k}(\mathbf{x},\mathbf{y})+\alpha_F(\mathbf{q})\right\}.\end{equation}
A minimizing distribution $\mathbf{q}_F$ in the penalty representation above supplies a supergradient of the follower utility and therefore yields the follower's strategy-optimality conditions.
\begin{proposition}[Convex-risk complementarity formulation]\label{prop:convex-risk-mpec}
Problem \eqref{eq:convex-risk-bilevel} is equivalent to
\begin{equation}\label{eq:convex-risk-mpec}
\begin{aligned}\max_{\mathclap{\mathbf{x},\mathbf{y},\mathbf{q}_F,\kappa_F}}\; &\quad U_L(\mathbf{x},\mathbf{y})\\ \mathrm{s.t.}\; &\quad \mathbf{x}\in\Delta(A_L),\; \mathbf{y}\geq\mathbf{0},\; \mathbf{1}^\top\mathbf{y}=1,\; 0\leq\kappa_F\leq1,\\ &\mathbf{q}_F\in\argmin_{\mathbf{q}\in\Delta_K}\Big\{\sum_{k=1}^Kq_kg_{F,k}(\mathbf{x},\mathbf{y})+\alpha_F(\mathbf{q})\Big\},\\ &0\leq\kappa_F-\sum_{k=1}^Kq_{F,k}\sum_{i\in A_L}x_i u_F(i,j\mid\xi_k)\perp y_j\geq0 &&j\in A_F.\end{aligned}
\end{equation}
\end{proposition}

Proposition~\ref{prop:convex-risk-mpec} is an exact characterization at the level of a general penalty. It becomes a MPEC when the constraint with the minimization over $\mathbf{q}$ is replaced by its optimality conditions. 
These are the KKT conditions of the convex Problem~\eqref{eq:follower-convex-risk-dual}. 

For entropic utility,
suppose player $\nu$ uses entropic utility with parameter $\theta_\nu>0$. 
By Equation~\eqref{eq:entropic-weights}, the follower's unique minimizing distribution has a closed form formula.
Substituting this into Proposition~\ref{prop:convex-risk-mpec} gives a formulation with no inner optimization problem where the $\argmin$ constraint changes to
$$
q_{F,k}=\frac{p_k\exp\bigl(-\theta_F g_{F,k}(\mathbf{x},\mathbf{y})\bigr)}{\sum_{k'=1}^K p_{k'} \exp\bigl(-\theta_F g_{F,k'}(\mathbf{x},\mathbf{y})\bigr)}, \quad k=1,\ldots,K.
$$
The resulting formulation is smooth apart from complementarity, but it is not generally convex. In Appendix~\ref{app:risk-aware-guards}, we show that the complementarity constraints can be replaced by an exact binary big-$M$ reformulation, yielding a mixed-integer nonlinear program for entropic utility that can be solved directly with Gurobi~\cite{gurobi}.

\paragraph{Gradient Method}\label{sec:implicit-gradient}

For this approach, we treat the follower response as an implicit function of the leader strategy $\mathbf x$ and applies projected gradient ascent to the resulting leader objective in order to obtain a scalable heuristic method for equilibrium finding. We keep assuming the conditions of Theorem~\ref{thm:convex-risk-attainment}, so the original follower response $\mathbf{y}^*(\mathbf{x})$ is unique and continuous. To apply implicit differentiation, we additionally assume that $U_F$ is twice continuously differentiable and that $U_L$ is continuously differentiable with respect to the players' mixed strategies. 
Since $U_F(\mathbf{x},\cdot)$ is concave, we add an action entropy term to the follower's payoff, to make it strictly concave and place its unique maximizer in the interior of the simplex. Together with the differentiability assumptions, these properties permit implicit differentiation of $\mathbf{y}_{\tau}(\mathbf{x})$. 

We first do action entropy regularization.
Define the Shannon entropy of the follower strategy by $\mathcal{H}(\mathbf{y})=-\sum_{j\in A_F}y_j\log y_j$ with $ 0\log0=0.$
Let the regularized follower response be
\begin{equation}\label{eq:entropy-regularized-follower-response}\mathbf{y}_{\tau}(\mathbf{x})=\argmax_{\mathbf{y}\in\Delta(A_F)}\left\{U_F(\mathbf{x},\mathbf{y})+\tau\mathcal{H}(\mathbf{y})\right\}.\end{equation}
The term $\tau\mathcal{H}(\mathbf{y})$ is action entropy, and is distinct from entropic risk in Equation~\eqref{eq:entropic-utility}. We use action entropy only as a computational regularizer and assume $\tau$ is close to zero. Accordingly, the final guarantee reported by the algorithm below is computed from the original utility $U_F$, without the action-entropy term. First, we bound the follower response approximation due to the action entropy.

\begin{proposition}[Follower approximation under action entropy]\label{prop:entropy-follower-approximation}
For every $\tau>0$ and $\mathbf{x}\in\Delta(A_L)$, the response $\mathbf{y}_{\tau}(\mathbf{x})$ in Equation~\eqref{eq:entropy-regularized-follower-response} exists and is unique. It is a $\tau\log n_F$-approximate response for the original follower problem:
\begin{equation}\label{eq:entropy-follower-gap}0\leq \max_{\mathbf{y}\in\Delta(A_F)}U_F(\mathbf{x},\mathbf{y})-U_F\!\left(\mathbf{x},\mathbf{y}_{\tau}(\mathbf{x})\right)\leq\tau\log n_F.\end{equation}
\end{proposition}

Thus, choosing $\tau\leq\epsilon_F/\log n_F$ makes the regularized response an $\epsilon_F$-response whenever $n_F\geq2$. Note that this statement is only for the follower's utility due to action entropy regularization. 

Next, we derive the implicit response derivative.
The action entropy term places $\mathbf{y}_{\tau}(\mathbf{x})$ in the interior of the follower simplex~\cite{sandholm2010population}, meaning that $y_{\tau,j}(\mathbf{x})>0$ for every $j\in A_F$. The assumed differentiability of $U_F$ then permits implicit differentiation at this response. 
Let $\kappa_{\tau}(\mathbf{x})$ be the multiplier of the follower probability sum equality. As nonnegativity constraints of probabilities are not active (interior solution), the first-order conditions for Equation~\eqref{eq:entropy-regularized-follower-response} are
\begin{equation}\label{eq:entropy-regularized-follower-kkt}\nabla_{\mathbf{y}}U_F(\mathbf{x},\mathbf{y}_{\tau})-\tau\bigl(\mathbf{1}+\log\mathbf{y}_{\tau}\bigr)-\kappa_{\tau}\mathbf{1}=\mathbf{0},\qquad \mathbf{1}^{\top}\mathbf{y}_{\tau}=1,\end{equation}
where the logarithm is componentwise. Define $\mathbf{B}_{\tau}(\mathbf{x})$ as the Hessian of the regularized follower objective in Equation~\eqref{eq:entropy-regularized-follower-response} with respect to $\mathbf{y}$, evaluated at $\mathbf{y}_{\tau}(\mathbf{x})$:
$
\mathbf{B}_{\tau}(\mathbf{x})
=\nabla^2_{\mathbf{y}\mathbf{y}}U_F\!\left(\mathbf{x},\mathbf{y}_{\tau}(\mathbf{x})\right)
-\tau\operatorname{Diag}\!\left(\mathbf{y}_{\tau}(\mathbf{x})\right)^{-1}.
$
The leader objective induced by the regularized response is
$
\Phi_{\tau}(\mathbf{x})=U_L\!\left(\mathbf{x},\mathbf{y}_{\tau}(\mathbf{x})\right).
$ Next, we provide the gradient of $\Phi_{\tau}$.

\begin{proposition}[Implicit follower and leader derivatives]\label{prop:implicit-gradient}
For every $\tau>0$, the regularized follower response $\mathbf{y}_{\tau}(\mathbf{x})$ is continuously differentiable with respect to the leader commitment $\mathbf{x}$. Here $D_{\mathbf{x}}\mathbf{y}_{\tau}(\mathbf{x})$ denotes its Jacobian with respect to $\mathbf{x}$.
\begin{equation}\label{eq:implicit-response-linear-system}\begin{bmatrix}\mathbf{B}_{\tau}(\mathbf{x})&-\mathbf{1}\\ \mathbf{1}^{\top}&0\end{bmatrix}\begin{bmatrix}D_{\mathbf{x}}\mathbf{y}_{\tau}(\mathbf{x})\\ D_{\mathbf{x}}\kappa_{\tau}(\mathbf{x})\end{bmatrix}=-\begin{bmatrix}\nabla^2_{\mathbf{y}\mathbf{x}}U_F\!\left(\mathbf{x},\mathbf{y}_{\tau}(\mathbf{x})\right)\\ \mathbf{0}^{\top}\end{bmatrix}.\end{equation}
Using $D_{\mathbf{x}}\mathbf{y}_{\tau}$ from above, the gradient of leader objective $\Phi_{\tau}(\mathbf{x})$ is
\begin{equation}\label{eq:smoothed-leader-implicit-gradient}
\nabla\Phi_{\tau}(\mathbf{x})
=\nabla_{\mathbf{x}}U_L\!\left(\mathbf{x},\mathbf{y}_{\tau}(\mathbf{x})\right)
+D_{\mathbf{x}}\mathbf{y}_{\tau}(\mathbf{x})^{\top}
\nabla_{\mathbf{y}}U_L\!\left(\mathbf{x},\mathbf{y}_{\tau}(\mathbf{x})\right).
\end{equation}
\end{proposition}

\paragraph{Projected Gradient with Random Restarts}
We use the gradient result in a projected gradient method shown in Algorithm~\ref{alg:implicit-gradient}.
The leader's probability simplex constraint is handled by Euclidean projection, denoted by $\Pi_{\Delta(A_L)}$. At an iterate $\mathbf{x}$, the algorithm solves the regularized follower problem, solves the linear system of equations \eqref{eq:implicit-response-linear-system}, and substitutes the result into the gradient in Equation~\eqref{eq:smoothed-leader-implicit-gradient}. Projected ascent with adaptive learning rate~\cite{ bertsekas1976goldstein} then increases $\Phi_{\tau}$. A candidate $\mathbf{z}$ is accepted if it satisfies the improvement condition in line~10; else, the current iterate $\mathbf x$ is retained and $\alpha$ is halved, while five accepted candidates in a row increase $\alpha$ by a factor of $1.2$. Next, we describe how the follower response problem is solved.

For fixed $\mathbf{x}$, we write $\mathbf{y}=\operatorname{softmax}(\mathbf{z})$ and solve the $n_F-1$ KKT equations $\partial_{y_j}U_F-\partial_{y_r}U_F-\tau\log(y_j/y_r)=0$, $j\neq r$, where $r$ is a reference action. We use Powell's hybrid method, which combines fast Newton steps with trust-region gradient steps when the Newton step is unreliable, together with a warm start from the preceding solution to the response problem~\cite{powell1970hybrid}. 

\begin{algorithm}[t]
\caption{Gradient method with adaptive learning rate}\label{alg:implicit-gradient}
\KwIn{Restarts $R$; continuation schedule of pairs $(\tau,T)$; tolerance $\zeta$; initial step size $\alpha_0$; $c\in(0,1)$}
\SetAlgoNoLine
\SetAlgoNoEnd
\For{$r\leftarrow1$ \KwTo $R$}{
Initialize $\mathbf{x}\in\Delta(A_L)$ and a follower warm start $\mathbf{y}$\;
\ForEach{$(\tau,T)$ in the continuation schedule}{
Solve \eqref{eq:entropy-regularized-follower-response} for $\mathbf{y}\leftarrow\mathbf{y}_{\tau}(\mathbf{x})$, warm-started from $\mathbf{y}$\;
\For{$t\leftarrow1$ \KwTo $T$}{
Compute $\mathbf{d}\leftarrow\nabla\Phi_{\tau}(\mathbf{x})$ using \eqref{eq:implicit-response-linear-system} and \eqref{eq:smoothed-leader-implicit-gradient}\;
\If{$\|\Pi_{\Delta(A_L)}(\mathbf{x}+\mathbf{d})-\mathbf{x}\|_2\leq\zeta$}{stop the current continuation stage\;}
$\mathbf{z}\leftarrow\Pi_{\Delta(A_L)}(\mathbf{x}+\alpha\mathbf{d})$,
solve $\overline{\mathbf{y}}\leftarrow\mathbf{y}_{\tau}(\mathbf{z})$\;
\uIf{$\Phi_{\tau}(\mathbf{z})\geq\Phi_{\tau}(\mathbf{x})+ c\mathbf{d}^{\top}(\mathbf{z}-\mathbf{x})$}{
 $(\mathbf{x},\mathbf{y})\leftarrow(\mathbf{z},\overline{\mathbf{y}})$; if this branch taken 5 times in a row: $\alpha\leftarrow\min\{1.2\alpha,\alpha_0\}$, else retain $\alpha$\;
}
\lElse{
Retain $(\mathbf{x},\mathbf{y})$ and set $\alpha \leftarrow \alpha/2$
}
}
}
$\mathbf{x}^r\leftarrow\mathbf{x}$, solve $\mathbf{y}^{*,r}\leftarrow\mathbf{y}^*(\mathbf{x}^r)$, and set $v_r\leftarrow U_L(\mathbf{x}^r,\mathbf{y}^{*,r})$\;
\Indp
}
Choose $r^*\in\argmax_{1\leq r\leq R}v_r$\;
\Return{$(\mathbf{x}^{r^*},\mathbf{y}^{*,r^*})$}\;
\end{algorithm}

With tolerance $\zeta$, the stopping rule measures first-order stationarity: $\|\Pi_{\Delta(A_L)}(\mathbf{x}+\nabla\Phi_{\tau}(\mathbf{x}))-\mathbf{x}\|_2$ is zero exactly when no feasible first-order change in the leader commitment increases $\Phi_{\tau}$. 
The next result assumes explicitly that at least one restart reaches a global maximizer.
Recall that $V=\max_{\mathbf{x}\in\Delta(A_L)}U_L(\mathbf{x},\mathbf{y}^*(\mathbf{x}))$ is the Stackelberg value defined in Equation~\eqref{eq:convex-risk-stackelberg-value}. We call $(\widehat{\mathbf{x}},\widehat{\mathbf{y}})$ an $\epsilon$-optimal Stackelberg solution if $\widehat{\mathbf{y}}=\mathbf{y}^*(\widehat{\mathbf{x}})$ and $U_L(\widehat{\mathbf{x}},\widehat{\mathbf{y}})\geq V-\epsilon$. 

\begin{theorem}\label{thm:implicit-gradient-approximation}
Fix $\tau>0$, and let $\mathbf{x}^{1},\ldots,\mathbf{x}^{R}$ be the strategies returned by the $R$ restarts. Suppose that, for some $g\in\{1,\ldots,R\}$, $\mathbf{x}^{g}$ is a global maximizer of $\Phi_{\tau}$. Algorithm~\ref{alg:implicit-gradient} selects
$
\widehat{\mathbf{x}}\in\argmax_{1\leq r\leq R}U_L\!\big(\mathbf{x}^{r},\mathbf{y}^*(\mathbf{x}^{r})\big)
$
and sets $\widehat{\mathbf{y}}=\mathbf{y}^*(\widehat{\mathbf{x}})$. Define $d_{\tau}=\max_{\mathbf{x}\in\Delta(A_L)}\|\mathbf{y}_{\tau}(\mathbf{x})-\mathbf{y}^*(\mathbf{x})\|_1$.
Then the solution returned satisfies
\begin{equation}\label{eq:implicit-gradient-general-guarantee}U_L(\widehat{\mathbf{x}},\widehat{\mathbf{y}})\geq V-2d_{\tau},\end{equation}
and $d_{\tau}\to0$ as $\tau \to0$. Consequently, for every $\epsilon>0$, sufficiently small $\tau$ makes the returned pair an $\epsilon$-Stackelberg equilibrium. If, in addition, the unregularized function $U_F(\mathbf{x},\cdot)$ is $\mu$-strongly concave in Euclidean norm with the same $\mu>0$ for every $\mathbf{x}$, then
\begin{equation}\label{eq:implicit-gradient-explicit-guarantee}U_L(\widehat{\mathbf{x}},\widehat{\mathbf{y}})\geq V-2\sqrt{ (2n_F\tau\log n_F)/\mu}.\end{equation}
For $n_F\geq2$, $\tau\leq\mu\epsilon^2/(8n_F\log n_F)$ suffices for an $\epsilon$-equilibrium.
\end{theorem}


\section{Experiments}\label{sec:experiments}

All our experiments were run on a MacBook Pro with a 10-core Apple M1 Max processor and 32 GB of memory, running macOS 26.6.2. We provide scalability and real-world qualitative experiments. 

\begin{figure*}
    \includegraphics[width=0.46\linewidth]{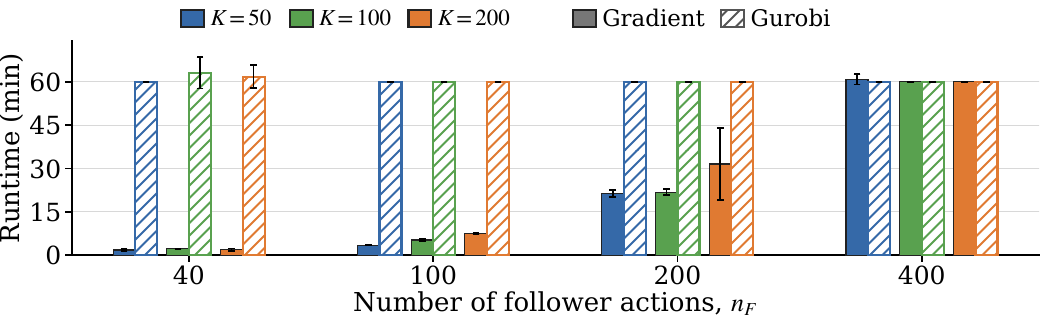}\hspace{10pt}
    \includegraphics[width=0.46\linewidth]{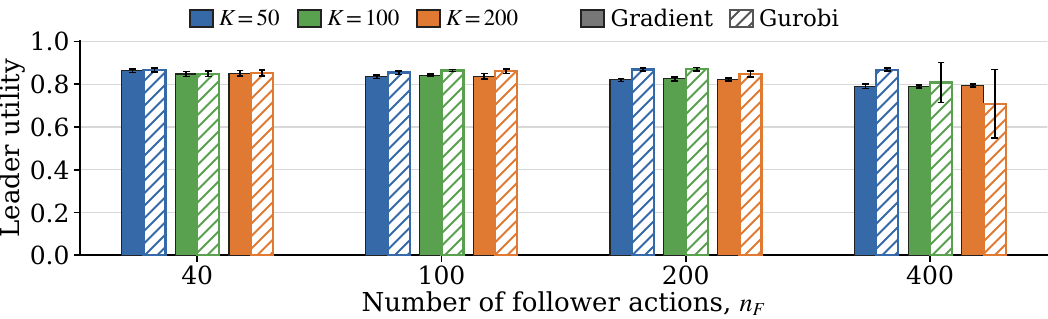}
    \caption{Runtime and solution quality for the scalability experiment for Gradient and Gurobi approaches} \label{fig:scalable}
\end{figure*}

\paragraph{Scalability} To evaluate scalability, we generate synthetic Stackelberg security games ~\cite{tambe2011} with $n_F\in\{40, 100, 200, 400\}$ targets and $K\in\{50,100,200\}$. Each of the $r$ identical defender resources covers one target; we set $r=n_F/2$, giving the deployment-to-saturation ratio $0.5$ identified as computationally challenging by \citet{jain2012deployment}. We generate payoff samples by perturbing target-specific \emph{defender losses} and \emph{attacker gains} for a successful attack with Gaussian shock, call these $d_{kt}$ and $a_{kt}$; the attacker always selects one target, and with an attack of target $t$ an uncovered (i.e., not defended) target has the defender–attacker rewards as $(1-d_{kt},a_{kt})$ that changes to $(1,0)$ when the target is covered.
We generate five game instances for each size combination and compare the \textbf{Gradient} method, using ten restarts, with exact MPEC solved by \textbf{Gurobi} under a one hour soft limit. We gradually decrease $\tau$ from $10^{-3}$ to $10^{-6}$ with total of 20,000 iterations, and use the adaptive learning rate as stated in algorithm initialized at $\alpha_0 = 10^{-2}$. More details are in Appendix~\ref{app:risk-aware-guards}.

The result in Figure~\ref{fig:scalable} shows that Gradient completes substantially earlier than Gurobi on small and medium instances, whereas both methods reach the one hour soft limit on the largest instances as repeatedly solving and implicitly differentiating through the follower response becomes costly. The methods achieve broadly comparable leader utilities; on the largest instances, neither uniformly dominates, while Gurobi exhibits substantially higher variability.

\begin{figure}[t]
    \centering
    \includegraphics[width=0.9\linewidth]{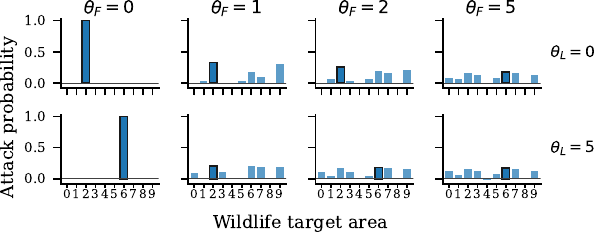}\\\vspace{4pt}
    \includegraphics[width=0.9\linewidth]{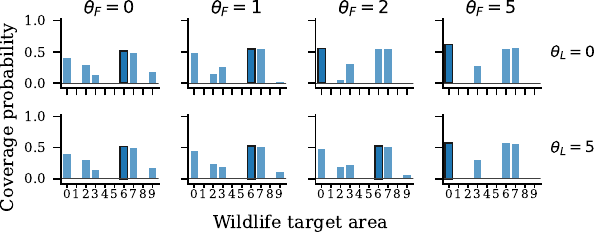}
    \caption{Attack and coverage prob. on the 10 targets with varying $\theta_F, \theta_L$ in simple schedule. Higher $\theta$: more risk-averse.}
    \label{fig:risk-aware-guards-target-profiles}
\end{figure}

\paragraph{\textsc{Risk-Aware-Guards}.} We construct \textsc{Risk-Aware-Guards} from the GUARD framework~\cite{krever2025guard}, which clusters 1,592 GPS-collar observations, spanning March 25, 2003 to May 8, 2007, from elephants in Lob\'ek\'e National Park into \emph{ten wildlife targets} and generates patrol schedules for \emph{two defender resources}. In the \emph{simple schedule}, each resource patrols one target at zero movement cost, whereas in the \emph{general schedule}, each resource may visit multiple targets, a joint patrol covers as many as seven distinct targets, and incurs route-dependent costs. After merging payoff-equivalent joint patrols and retaining the minimum-cost representative for each coverage pattern, the simple and general instances contain 55 and 401 effective joint patrol actions, respectively, and ten attacker actions. We model an environmental realization using one of the elephant location observations: state $\xi_k$ occurs when the observation belongs to target $k$ and therefore has empirical probability $p_k=N_k/1592$, where $N_k$ is the number of observations in that target. The $\xi_k$-dependent payoffs are constructed so that their expectations reproduce GUARD's payoffs for every joint patrol and attacked target, and both players evaluate the resulting random rewards using entropic utilities with parameters $\theta_L,\theta_F$ ($\theta=0$ is risk-neutral). Appendix~\ref{app:risk-aware-guards} gives the complete data, payoff, and preprocessing construction.

\begin{figure}[t]
    \centering
    \includegraphics[width=0.94\linewidth]{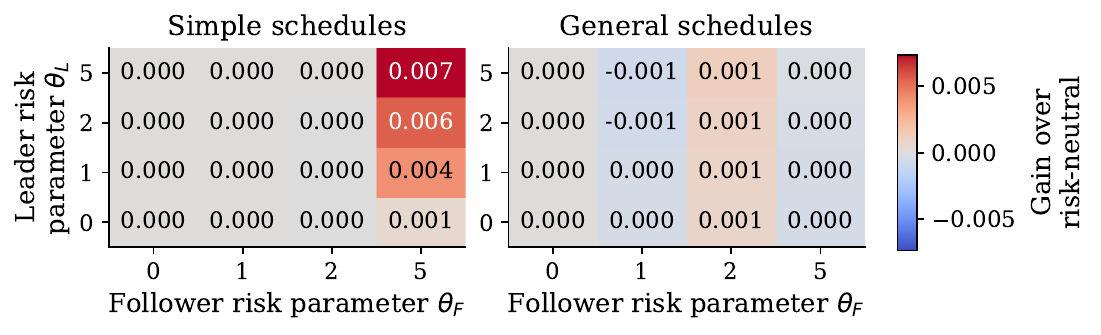}
    \caption{Gain over risk-neutral defender strategy ($\theta_L\!=\!\theta_F\!=\!0$) . Each cell reports $U_L^{\theta_L,\mathrm{test}}\!(\mathbf{x}^{\mathrm{train}}_{\theta_L,\theta_F},\mathbf{y}^{*,\mathrm{test}}_{\theta_F}(\mathbf{x}^{\mathrm{train}}_{\theta_L,\theta_F}))-U_L^{\theta_L,\mathrm{test}}\!(\mathbf{x}^{\mathrm{train}}_{0,0},\mathbf{y}^{*,\mathrm{test}}_{\theta_F}(\mathbf{x}^{\mathrm{train}}_{0,0}))$; follower reoptimizes at test time.}
    \label{fig:gain}
\end{figure}

Figure~\ref{fig:risk-aware-guards-target-profiles} shows that increasing $\theta_F$ disperses the attacker’s response while concentrating coverage on high-frequency wildlife targets; $\theta_L$ primarily affects the magnitude of this reallocation. The same qualitative pattern holds for the omitted leader parameters and for the risk-sensitive cases in general schedule (Appendix~\ref{app:addtional}). Next, we perform a chronological 70:30 train-test split and evaluate the defender strategy from training data after allowing the attacker to reoptimize on the test distribution. Figure~\ref{fig:gain} (see caption) shows that risk-aware training produces essentially no loss in held-out defender utility relative to the $(0,0)$-trained commitment. Gains are most pronounced under simple schedules for $\theta_F=5$; under general schedules they remain near 0, consistent with the greater coverage flexibility of joint patrols that can cover up to 7 targets.

\section{Conclusion}
We introduced, to our knowledge, the first systematic treatment of Stackelberg games in which both players evaluate environment-induced random rewards using convex-risk utilities. We prove NP-hardness and develop two computational methods: for entropic utility, we obtain an optimization formulation amenable to high-performance solvers such as Gurobi, while the gradient method applies more broadly whenever the smooth convex-risk follower response can be computed and differentiated. Extending this framework to multiple followers or defenders with heterogeneous risk preferences is an important direction for future work.

\newpage
\clearpage
\balance
\bibliographystyle{ACM-Reference-Format}
\bibliography{references}

\newpage 

\appendix

\section{Missing Proofs}
\begin{proof}[\textbf{Proof of Proposition~\ref{prop:general-cusg-existence}}]
Under the finite-support representation, each $\rho_\nu$ is a finite-valued concave function on the finite-dimensional space $\mathbb{R}^K$ and is therefore continuous. Since every $g_{\nu,k}(\mathbf{x},\mathbf{y})$ is bilinear in $(\mathbf{x},\mathbf{y})$, each utility $U_\nu$ is continuous on the compact set $\Delta(A_L)\times\Delta(A_F)$.

Fix $\mathbf{x}\in\Delta(A_L)$. Continuity of $U_F(\mathbf{x},\cdot)$ and compactness of $\Delta(A_F)$ imply that $BR_F(\mathbf{x})$ is nonempty and compact. We next show that the graph of $BR_F$ is closed. Consider any sequences $\mathbf{x}^{n}\to\mathbf{x}$ and $\mathbf{y}^{n}\to\mathbf{y}$ satisfying $\mathbf{y}^{n}\in BR_F(\mathbf{x}^{n})$. For every $\widetilde{\mathbf{y}}\in\Delta(A_F)$, follower optimality gives
$
U_F(\mathbf{x}^{n},\mathbf{y}^{n})\geq U_F(\mathbf{x}^{n},\widetilde{\mathbf{y}}).$
Taking $n\to\infty$ and using continuity yields $U_F(\mathbf{x},\mathbf{y})\geq U_F(\mathbf{x},\widetilde{\mathbf{y}})$. Since this holds for every $\widetilde{\mathbf{y}}$, we have $\mathbf{y}\in BR_F(\mathbf{x})$. Hence
\begin{equation}\label{eq:follower-response-graph}\operatorname{Gr}(BR_F)=\left\{(\mathbf{x},\mathbf{y})\in\Delta(A_L)\times\Delta(A_F):\mathbf{y}\in BR_F(\mathbf{x})\right\}\end{equation}
is closed. It is a closed subset of a compact set and is therefore compact. Continuity of $U_L$ now implies that it attains a maximum on $\operatorname{Gr}(BR_F)$. This maximum is precisely $V_L^+$, proving existence of an optimistic Stackelberg equilibrium.

To prove the second claim, consider two leader actions $A_L=\{L_1,L_2\}$ and two follower actions $A_F=\{a,b\}$. Write $\mathbf{x}=(p,1-p)$ and $\mathbf{y}=(q,1-q)$, where $q$ is the probability of action $a$. In every realization $\xi_k$, let the follower and leader payoff matrices be
\begin{equation}\label{eq:pessimistic-nonattainment-matrices}\bigl(u_F(i,j\mid\xi_k)\bigr)_{ij}=\begin{pmatrix}0&1\\1&0\end{pmatrix},\qquad \bigl(u_L(i,j\mid\xi_k)\bigr)_{ij}=\begin{pmatrix}1&0\\0&0\end{pmatrix}.\end{equation}
The resulting rewards are deterministic across realizations. Definition~\ref{def:convex-monetary-utility} implies $\rho_\nu(c)=\rho_\nu(0+c)=\rho_\nu(0)+c=c$ for every constant $c$. Therefore,
\begin{equation}\label{eq:pessimistic-nonattainment-utilities}U_F(p,q)=p+q-2pq,\qquad U_L(p,q)=pq.\end{equation}
For fixed $p$, the coefficient of $q$ in $U_F(p,q)$ is $1-2p$. Consequently,
\begin{equation}\label{eq:pessimistic-nonattainment-response}BR_F(p)=\begin{cases}\{1\},&0\leq p<\frac{1}{2},\\[2pt][0,1],&p=\frac{1}{2},\\[2pt]\{0\},&\frac{1}{2}<p\leq1.\end{cases}\end{equation}
The pessimistic leader utility induced by commitment $p$ is therefore
\begin{equation}\label{eq:pessimistic-nonattainment-objective}\min_{q\in BR_F(p)}U_L(p,q)=\begin{cases}p,&0\leq p<\frac{1}{2},\\[2pt]0,&\frac{1}{2}\leq p\leq1.\end{cases}\end{equation}
Its supremum is $1/2$, because values arbitrarily close to $1/2$ are obtained as $p$ approaches $1/2$ from below. No commitment attains this value: the objective is strictly below $1/2$ when $p<1/2$ and equals zero when $p\geq1/2$. Hence a pessimistic Stackelberg equilibrium need not exist.
\end{proof}

\begin{proof}[\textbf{Proof of Theorem~\ref{thm:convex-risk-attainment}}]
Fix $\mathbf{x}\in\Delta(A_L)$. The function $U_F(\mathbf{x},\cdot)$ is continuous on the compact simplex $\Delta(A_F)$, so it has at least one maximizer. 
{\color{black}We first show that the dual optimizer $\mathbf{q}_F$ supplied by Assumption~\ref{ass:active-tie-independence} supports every follower best response at this fixed $\mathbf{x}$.} Define
\begin{equation}\label{eq:follower-penalty-saddle-function}L_{\mathbf{x}}(\mathbf{y},\mathbf{q})=\sum_{k=1}^Kq_k g_{F,k}(\mathbf{x},\mathbf{y})+\alpha_F(\mathbf{q}).\end{equation}
By Equation~\eqref{eq:convex-penalty-representation}, minimizing $L_{\mathbf{x}}(\mathbf{y},\cdot)$ yields $U_F(\mathbf{x},\mathbf{y})$. Because $\min_{\mathbf{q}}\alpha_F(\mathbf{q})=0$, choose $\mathbf{q}^0$ with $\alpha_F(\mathbf{q}^0)=0$. Since $0\leq g_{F,k}\leq1$, we have $L_{\mathbf{x}}(\mathbf{y},\mathbf{q}^0)\leq1$ for every $\mathbf{y}$. Conversely, if $\alpha_F(\mathbf{q})>1$, then $L_{\mathbf{x}}(\mathbf{y},\mathbf{q})\geq\alpha_F(\mathbf{q})>1$ for every $\mathbf{y}$. Such a distribution cannot minimize either the inner problem or its maximum over $\mathbf{y}$. We may therefore restrict $\mathbf{q}$ to the nonempty compact convex sublevel set $\mathcal{Q}_1=\{\mathbf{q}\in\Delta_K:\alpha_F(\mathbf{q})\leq1\}$, which allows applying the minimax theorem next.

On $\Delta(A_F)\times\mathcal{Q}_1$, the function $L_{\mathbf{x}}$ is affine and continuous in $\mathbf{y}$ and convex and lower semicontinuous in $\mathbf{q}$. The finite-dimensional minimax theorem therefore gives
\begin{equation}\label{eq:general-penalty-minimax}\max_{\mathbf{y}\in\Delta(A_F)}\min_{\mathbf{q}\in\mathcal{Q}_1}L_{\mathbf{x}}(\mathbf{y},\mathbf{q})=\min_{\mathbf{q}\in\mathcal{Q}_1}\max_{\mathbf{y}\in\Delta(A_F)}L_{\mathbf{x}}(\mathbf{y},\mathbf{q})=:M.\end{equation}
{\color{black}Choose the minimizer $\mathbf{q}_F$ of Problem~\eqref{eq:follower-minimax-dual} for which Assumption~\ref{ass:active-tie-independence} guarantees affine independence. Since $L_{\mathbf{x}}$ is affine in the follower strategy, $\max_{\widetilde{\mathbf{y}}\in\Delta(A_F)}L_{\mathbf{x}}(\widetilde{\mathbf{y}},\mathbf{q})=\alpha_F(\mathbf{q})+\max_{j\in A_F}\sum_kq_k g_{F,k}(\mathbf{x},\mathbf{e}_j)$. The restriction argument above shows that $\mathbf{q}_F\in\mathcal{Q}_1$, so it minimizes the right side of Equation~\eqref{eq:general-penalty-minimax}. For any $\mathbf{y}\in BR_F(\mathbf{x})$, follower optimality gives $M=\min_{\mathbf{q}\in\mathcal{Q}_1}L_{\mathbf{x}}(\mathbf{y},\mathbf{q})\leq L_{\mathbf{x}}(\mathbf{y},\mathbf{q}_F)$, while dual optimality gives $L_{\mathbf{x}}(\mathbf{y},\mathbf{q}_F)\leq\max_{\widetilde{\mathbf{y}}\in\Delta(A_F)}L_{\mathbf{x}}(\widetilde{\mathbf{y}},\mathbf{q}_F)=M$. Therefore $L_{\mathbf{x}}(\mathbf{y},\mathbf{q}_F)=M$, and every such $\mathbf{y}$ maximizes $L_{\mathbf{x}}(\cdot,\mathbf{q}_F)$.\par}

The penalty term is constant in $\mathbf{y}$, so $\mathbf{y}$ maximizes the linear function $\sum_{j\in A_F}y_j\sum_{k=1}^Kq_{F,k}g_{F,k}(\mathbf{x},\mathbf{e}_j)$. It follows that
\begin{equation}\label{eq:follower-support-active-tie}\operatorname{supp}(\mathbf{y})\subseteq J_F(\mathbf{x},\mathbf{q}_F).\end{equation}

Now, suppose $\mathbf{y},\mathbf{y}'\in BR_F(\mathbf{x})$. Let $X=g_F(\mathbf{x},\mathbf{y}\mid\xi)$ and $Y=g_F(\mathbf{x},\mathbf{y}'\mid\xi)$. If $X-Y$ were not almost surely constant, strict concavity modulo translations and the affinity of the mixed payoff would give
\begin{equation}\label{eq:two-optima-strict-curvature}U_F\!\left(\mathbf{x},\frac{\mathbf{y}+\mathbf{y}'}{2}\right)>\frac{1}{2}U_F(\mathbf{x},\mathbf{y})+\frac{1}{2}U_F(\mathbf{x},\mathbf{y}')=M,\end{equation}
contradicting the definition of $M$. Hence $X-Y=c$ almost surely for some $c\in\mathbb{R}$. Translation equivariance and optimality give $M=\rho_F(X)=\rho_F(Y+c)=\rho_F(Y)+c=M+c$, so $c=0$. Thus $X=Y$ almost surely and, because every empirical probability is positive,
\begin{equation}\label{eq:two-optima-same-payoff-vector}\mathbf{g}_F(\mathbf{x},\mathbf{y})=\mathbf{g}_F(\mathbf{x},\mathbf{y}').\end{equation}
{\color{black}Equation~\eqref{eq:two-optima-same-payoff-vector} gives $L_{\mathbf{x}}(\mathbf{y}',\mathbf{q}_F)=L_{\mathbf{x}}(\mathbf{y},\mathbf{q}_F)=M$. Dual optimality gives $\max_{\widetilde{\mathbf{y}}\in\Delta(A_F)}L_{\mathbf{x}}(\widetilde{\mathbf{y}},\mathbf{q}_F)=M$, so $\mathbf{y}'$ also maximizes $L_{\mathbf{x}}(\cdot,\mathbf{q}_F)$. Since the penalty $\alpha_F(\mathbf{q}_F)$ is independent of the follower strategy, the same linear-maximization argument leading to Equation~\eqref{eq:follower-support-active-tie} gives $\operatorname{supp}(\mathbf{y}')\subseteq J_F(\mathbf{x},\mathbf{q}_F)$. Both $\mathbf{y}$ and $\mathbf{y}'$ satisfy Equation~\eqref{eq:follower-support-active-tie} for the same $\mathbf{q}_F$ selected using Assumption~\ref{ass:active-tie-independence}. Combining their simplex equalities with Equation~\eqref{eq:two-optima-same-payoff-vector} gives
\begin{equation}\label{eq:two-optima-affine-dependence}\sum_{j\in J_F(\mathbf{x},\mathbf{q}_F)}(y_j-y_j')=0,\quad \sum_{j\in J_F(\mathbf{x},\mathbf{q}_F)}(y_j-y_j')\mathbf{g}_F(\mathbf{x},\mathbf{e}_j)=\mathbf{0}.\end{equation}
The vectors indexed by this selected tied set are affinely independent, so every coefficient $y_j-y_j'$ is zero. Both strategies vanish outside that set; hence $\mathbf{y}=\mathbf{y}'$, proving uniqueness.\par}

The same argument gives the support bound. Every vector indexed by $J_F(\mathbf{x},\mathbf{q}_F)$ satisfies
\begin{equation}\label{eq:active-tie-supporting-hyperplane}\sum_{k=1}^K q_{F,k}g_{F,k}(\mathbf{x},\mathbf{e}_j)=\max_{j'\in A_F}\sum_{k=1}^Kq_{F,k}g_{F,k}(\mathbf{x},\mathbf{e}_{j'}),\; j\in J_F(\mathbf{x},\mathbf{q}_F).\end{equation}
Let
$
c=\max_{j'\in A_F}\sum_{k=1}^Kq_{F,k}g_{F,k}(\mathbf{x},\mathbf{e}_{j'}).
$
Equation~\eqref{eq:active-tie-supporting-hyperplane} implies that, for every $j\in J_F(\mathbf{x},\mathbf{q}_F)$,
$
\sum_{k=1}^Kq_{F,k}g_{F,k}(\mathbf{x},\mathbf{e}_j)=c.
$
{\color{black}Thus, every vector $\mathbf{g}_F(\mathbf{x},\mathbf{e}_j)$ indexed by $J_F(\mathbf{x},\mathbf{q}_F)$ belongs to the set $\big\{\mathbf{z}\in\mathbb{R}^K:\sum_{k=1}^Kq_{F,k}z_k=c\big\}$.}
Because $\mathbf{q}_F\in\Delta_K$, we have $\mathbf{q}_F\neq\mathbf{0}$, so this set is defined by one nontrivial affine equality in $\mathbb{R}^K$ and is therefore a hyperplane of dimension $K-1$. 
{\color{black}By the choice of $\mathbf{q}_F$ in Assumption~\ref{ass:active-tie-independence}, the vectors in this particular tied set are affinely independent.} An affinely independent subset of a $(K-1)$-dimensional affine space contains at most $K$ vectors. Combining this observation with Equation~\eqref{eq:follower-support-active-tie} therefore gives $|\operatorname{supp}(\mathbf{y}^*(\mathbf{x}))|\leq\min\{n_F,K\}$.

To prove continuity, take any sequence $\mathbf{x}^r\to\mathbf{x}$ and write $\mathbf{y}^r=\mathbf{y}^*(\mathbf{x}^r)$. Compactness of $\Delta(A_F)$ implies that every subsequence of $(\mathbf{y}^r)$ has a further subsequence converging to some $\overline{\mathbf{y}}$. Follower optimality gives $U_F(\mathbf{x}^r,\mathbf{y}^r)\geq U_F(\mathbf{x}^r,\mathbf{y})$ for every $\mathbf{y}\in\Delta(A_F)$. Taking limits along the convergent subsequence and using continuity of $U_F$ yields $U_F(\mathbf{x},\overline{\mathbf{y}})\geq U_F(\mathbf{x},\mathbf{y})$ for every $\mathbf{y}$. Hence $\overline{\mathbf{y}}=\mathbf{y}^*(\mathbf{x})$ by uniqueness. Every convergent subsequence has the same limit, so $\mathbf{y}^r\to\mathbf{y}^*(\mathbf{x})$ and the response map is continuous.

Finally, $U_L$ is continuous and therefore $\mathbf{x}\mapsto U_L(\mathbf{x},\mathbf{y}^*(\mathbf{x}))$ is continuous. It attains its maximum on the compact leader simplex. Since the follower response set contains exactly one strategy at every commitment, favorable and unfavorable tie-breaking select the same response, proving the last claim.
\end{proof}

\begin{proof}[\textbf{Proof of Lemma~\ref{lem:entropic-properties}}]
Let $\mathbf z=(X(\xi_1),\ldots,X(\xi_K))$ and let $\mathbf q(\mathbf z)$ denote the entropic weights from \eqref{eq:entropic-weights}. Since $p_k>0$ for every $k$, we have $q_k(\mathbf z)>0$ for every $k$. Differentiating the entropic utility with respect to the sample payoff coordinates gives $\partial\rho_\theta(\mathbf z)/\partial z_k=q_k(\mathbf z)$, and hence
\[
\nabla^2\rho_\theta(\mathbf z)=-\theta\left(\operatorname{Diag}(\mathbf q(\mathbf z))-\mathbf q(\mathbf z)\mathbf q(\mathbf z)^\top\right).
\]
Therefore, for any direction $\mathbf d\in\mathbb R^K$,
\[
\mathbf d^\top\nabla^2\rho_\theta(\mathbf z)\mathbf d=-\theta\left(\sum_{k=1}^Kq_k(\mathbf z)d_k^2-\left(\sum_{k=1}^Kq_k(\mathbf z)d_k\right)^2\right).
\]
The term in parentheses is the variance of the values $d_1,\ldots,d_K$ under the strictly positive distribution $\mathbf q(\mathbf z)$. It is zero if and only if $d_1=\cdots=d_K$, equivalently, $\mathbf d$ is proportional to $\mathbf 1$. Thus the quadratic form is strictly negative for every direction not proportional to $\mathbf 1$.

Now let $\mathbf d=(X(\xi_1)-Y(\xi_1),\ldots,X(\xi_K)-Y(\xi_K))$. If $X-Y$ is not constant across the samples, then $\mathbf d$ is not proportional to $\mathbf 1$. Hence the function $f(t)=\rho_\theta(Y+t(X-Y))$ satisfies $f''(t)<0$ for every $t\in(0,1)$. Therefore $f$ is strictly concave, and for every $\lambda\in(0,1)$,
\[
\rho_\theta(\lambda X+(1-\lambda)Y)>\lambda\rho_\theta(X)+(1-\lambda)\rho_\theta(Y).
\]
Hence $\rho_\theta$ is strictly concave modulo translations.
\end{proof}

\begin{proof}[\textbf{Proof of Proposition~\ref{prop:convex-risk-mpec}}]
Fix $\mathbf{x}$ and $\mathbf{y}$ and let $\mathbf{q}_F$ attain the minimum in Equation~\eqref{eq:follower-convex-risk-dual}. Since Equation~\eqref{eq:convex-penalty-representation} is the pointwise minimum of affine functions of the payoff vector, $\mathbf{q}_F$ is a supergradient of $\rho_F$ at $\mathbf{g}_F(\mathbf{x},\mathbf{y})$. Because $g_{F,k}(\mathbf{x},\mathbf{y})$ is affine in $\mathbf{y}$, the corresponding supergradient component with respect to $y_j$ is
\begin{equation}\label{eq:convex-risk-follower-marginal}\sum_{k=1}^Kq_{F,k}\frac{\partial g_{F,k}(\mathbf{x},\mathbf{y})}{\partial y_j}=\sum_{k=1}^Kq_{F,k}\sum_{i\in A_L}x_i u_F(i,j\mid\xi_k).\end{equation}
If the dual minimizer is not unique, the minimizing set is convex and the expression in Equation~\eqref{eq:convex-risk-follower-marginal} is linear in $\mathbf{q}_F$. Hence a suitable minimizing distribution represents any convex combination of active supergradients required by the follower optimality conditions.

The follower maximizes a concave function over the simplex. Its KKT conditions are therefore necessary and sufficient: there must initially be an unrestricted multiplier $\kappa_F$ for $\mathbf{1}^\top\mathbf{y}=1$ such that every marginal in Equation~\eqref{eq:convex-risk-follower-marginal} is at most $\kappa_F$, with equality whenever $y_j>0$. At least one coordinate of $\mathbf{y}$ is positive, so $\kappa_F$ equals the marginal of a used action. That marginal is a convex combination, first over leader actions and then over environmental samples, of numbers in $[0,1]$. Hence $0\leq\kappa_F\leq1$, and these conditions are exactly the last line of Problem~\eqref{eq:convex-risk-mpec} together with the stated bound.

To verify sufficiency directly, take any $\widetilde{\mathbf{y}}\in\Delta(A_F)$. The supergradient inequality gives
\[
U_F(\mathbf{x},\widetilde{\mathbf{y}})\leq U_F(\mathbf{x},\mathbf{y})+\sum_{j\in A_F}\left(\sum_{k=1}^Kq_{F,k}\sum_{i\in A_L}x_i u_F(i,j\mid\xi_k)\right)(\widetilde y_j-y_j).
\]
The complementarity constraints imply that the coefficient of every $\widetilde y_j$ is at most $\kappa_F$ and that the coefficient of every positive $y_j$ equals $\kappa_F$. Since both strategies sum to one, the final term is at most $\kappa_F-\kappa_F=0$. Thus $\mathbf{y}$ is a follower best response. Conversely, the concave KKT conditions at any follower best response provide a compatible active supergradient, minimizing distribution, and multiplier. The feasible pairs in Equation~\eqref{eq:convex-risk-mpec} are therefore exactly the feasible leader--follower pairs in Equation~\eqref{eq:convex-risk-bilevel}, and maximizing $U_L$ implements optimistic tie-breaking.
\end{proof}

\begin{proof}[\textbf{Proof of Lemma~\ref{lem:continuous-cubic-max-cut}}]
Fix any feasible $\mathbf{t}$ and clip each coordinate at one by setting $\bar t_i=\min\{t_i,1\}$. For every coordinate, $t_i-\bar t_i=(t_i-1)_+$: both sides are zero when $t_i\leq1$, while both equal $t_i-1$ when $t_i>1$. Hence, for every edge $\{i,j\}\in E$, we can write $t_i-t_j=(\bar t_i-\bar t_j)+((t_i-1)_+-(t_j-1)_+)$. Applying $|a+b|\leq|a|+|b|$ and then $|r-s|\leq r+s$ for the nonnegative numbers $r=(t_i-1)_+$ and $s=(t_j-1)_+$ gives
\begin{equation*}\label{eq:cut-clipping-edge-bound}|t_i-t_j|\leq|\bar t_i-\bar t_j|+(t_i-1)_++(t_j-1)_+.\end{equation*}
Summing the above over all edges counts $(t_i-1)_+$ exactly three times because every vertex has degree three. Consequently,
\begin{equation}\label{eq:cut-clipping-sum-bound}S_G(\mathbf{t})\leq S_G(\bar{\mathbf{t}})+3P(\mathbf{t}),\end{equation}
and hence $S_G(\mathbf{t})-4P(\mathbf{t})\leq S_G(\bar{\mathbf{t}})-P(\mathbf{t})\leq S_G(\bar{\mathbf{t}})$.

It remains to bound the continuous cut value of $\bar{\mathbf{t}}\in[0,1]^n$. For every threshold $z\in[0,1]$, let $C_z=\{i\in V:\bar t_i\geq z\}$. For each edge $\{i,j\}$, the indicator that $C_z$ separates its endpoints equals one precisely for thresholds lying between $\bar t_i$ and $\bar t_j$. Therefore,
\begin{equation}\label{eq:cut-threshold-integral}S_G(\bar{\mathbf{t}})=\int_0^1\left|\bigl\{\{i,j\}\in E:i\in C_z,\ j\notin C_z\bigr\}\right|\,\mathrm{d}z\leq\operatorname{MC}(G).\end{equation}
Equations~\eqref{eq:cut-clipping-sum-bound} and \eqref{eq:cut-threshold-integral} show that the left side of Equation~\eqref{eq:continuous-cubic-max-cut-identity} is at most $\operatorname{MC}(G)$. For the reverse inequality, let $C^*$ be a maximum cut and set $t_i=1$ when $i\in C^*$ and $t_i=0$ otherwise. This vector satisfies $\mathbf{1}^{\top}\mathbf{t}\leq n$, has $P(\mathbf{t})=0$, and gives $S_G(\mathbf{t})=\operatorname{MC}(G)$. Thus equality holds.
\end{proof}

\begin{proof}[\textbf{Proof of Proposition~\ref{prop:additive-hardness-threshold-response}}]
Let $\mu$ be the multiplier for $\sum_s y_s=1$ and let $\eta_s\geq0$ be the multiplier for $y_s\geq0$. The stationarity and complementary-slackness conditions are
\begin{equation}\label{eq:additive-hardness-follower-kkt}-a\exp\bigl(d_s(\mathbf{x})-ay_s\bigr)+\mu-\eta_s=0,\qquad \eta_sy_s=0,\qquad s\in A_F.\end{equation}
At least one coordinate of $\mathbf{y}$ is positive. For any $s$ with $y_s>0$, complementary slackness gives $\eta_s=0$, and hence $\mu=a\exp(d_s(\mathbf{x})-ay_s)>0$. Define $\tau(\mathbf{x})=\log(\mu/a)$. If $y_s>0$, stationarity gives $d_s(\mathbf{x})-ay_s=\tau(\mathbf{x})$. If $y_s=0$, stationarity gives $\eta_s=\mu-a\exp(d_s(\mathbf{x}))\geq0$, which is equivalent to $d_s(\mathbf{x})\leq\tau(\mathbf{x})$. Combining the two cases gives the first relation in Equation~\eqref{eq:additive-hardness-threshold-response}; summing over $s$ and using $\sum_s y_s=1$ gives the second.

For fixed $\mathbf{x}$, the function $\tau\mapsto\sum_s(d_s(\mathbf{x})-\tau)_+$ is continuous and strictly decreasing whenever it is positive, tends to infinity as $\tau\to-\infty$, and equals zero for $\tau\geq\max_s d_s(\mathbf{x})$. Therefore, the second relation in Equation~\eqref{eq:additive-hardness-threshold-response} determines a unique threshold. Strict convexity of the follower objective then gives the unique response $\mathbf{y}^*(\mathbf{x})$.

It remains to bound the threshold. For each edge $\{i,j\}\in E$, Equation~\eqref{eq:additive-hardness-feature-scores} and $(z)_++(-z)_+=|z|$ give
\begin{equation*}\bar y_{e_{ij}}(\mathbf{x})+\bar y_{e_{ji}}(\mathbf{x})=\frac{\lambda}{m}|t_i-t_j|.\end{equation*}
Each node action contributes $\bar y_{p_i}(\mathbf{x})=4\lambda(t_i-1)_+/m$, while every anchor contributes zero. Consequently,
\begin{equation}\label{eq:additive-hardness-zero-threshold-mass}\sum_{s\in A_F}\bar y_s(\mathbf{x})=\frac{1}{a}\sum_{s\in A_F}(d_s(\mathbf{x}))_+=\frac{\lambda}{m}\left(S_G(\mathbf{t})+4P(\mathbf{t})\right).\end{equation}
Since $\mathbf{t}\geq\mathbf{0}$, $\mathbf{1}^{\top}\mathbf{t}\leq n$, and $G$ is $3$-regular,
\begin{equation}\label{eq:additive-hardness-feature-mass-bound}S_G(\mathbf{t})\leq\sum_{\{i,j\}\in E}(t_i+t_j)=3\sum_{i=1}^nt_i\leq3n,\qquad P(\mathbf{t})\leq\sum_{i=1}^nt_i\leq n.\end{equation}
Using $m=3n/2$ and $\lambda=1/16$ in Equations~\eqref{eq:additive-hardness-zero-threshold-mass} and \eqref{eq:additive-hardness-feature-mass-bound} yields
\begin{equation*}\sum_{s\in A_F}\bar y_s(\mathbf{x})\leq\frac{14\lambda}{3}=\frac{7}{24}<1.\end{equation*}
If $\tau(\mathbf{x})\geq0$, Equation~\eqref{eq:additive-hardness-threshold-response} would imply
\begin{equation*}1=\frac{1}{a}\sum_s(d_s(\mathbf{x})-\tau(\mathbf{x}))_+\leq\frac{1}{a}\sum_s(d_s(\mathbf{x}))_+=\sum_s\bar y_s(\mathbf{x})<1,\end{equation*}
which is impossible. Hence $\tau(\mathbf{x})<0$. Since $d_{z_r}(\mathbf{x})=0$, every anchor receives probability $y_{z_r}^*(\mathbf{x})=-\tau(\mathbf{x})/a=e(\mathbf{x})>0$. Their total probability is $Me(\mathbf{x})\leq1$, proving $e(\mathbf{x})\leq (1/M)$. Finally, Equation~\eqref{eq:additive-hardness-threshold-response} gives the first relation in Equation~\eqref{eq:additive-hardness-feature-error}; the second follows from $0\leq(v+e)_+-(v)_+\leq e$ for every $v\in\mathbb{R}$ and $e\geq0$.
\end{proof}

\begin{proof}[\textbf{Proof of Proposition~\ref{prop:additive-hardness-value-encoding}}]
The payoffs in Equation~\eqref{eq:additive-hardness-leader-payoffs} lie in $[0,1]$ and are independent of the leader action and environmental sample. Therefore, every sample gives the leader reward $Y_E(\mathbf{x})+Y_0(\mathbf{x})/2$. Since $Y_E(\mathbf{x})+Y_P(\mathbf{x})+Y_0(\mathbf{x})=1$, this reward equals $1/2+(Y_E(\mathbf{x})-Y_P(\mathbf{x}))/2$. Entropic utility evaluates a reward that is constant across all samples at that same reward, proving Equation~\eqref{eq:additive-hardness-leader-deterministic-value}.

The edge calculation in the proof of Proposition~\ref{prop:additive-hardness-threshold-response} gives $\bar Y_E(\mathbf{x})=\lambda S_G(\mathbf{t})/m$, while the node calculation gives $\bar Y_P(\mathbf{x})=4\lambda P(\mathbf{t})/m$. Substitution into the definition of $\bar W$ proves $\bar Y_E(\mathbf{x})=\frac{\lambda}{m}S_G(\mathbf{t})$, $ \bar Y_P(\mathbf{x})=\frac{4\lambda}{m}P(\mathbf{t})$, $\bar W(\mathbf{x})=\frac{1}{2}+\frac{\lambda}{2m}\left(S_G(\mathbf{t})-4P(\mathbf{t})\right)$. The leader simplex represents exactly the set $\{\mathbf{t}\geq\mathbf{0}:\mathbf{1}^{\top}\mathbf{t}\leq n\}$, so Lemma~\ref{lem:continuous-cubic-max-cut} gives Equation~\eqref{eq:additive-hardness-intended-optimum}.

Finally, Proposition~\ref{prop:additive-hardness-threshold-response} gives $|y_s^*(\mathbf{x})-\bar y_s(\mathbf{x})|\leq1/M$ for every non-anchor action. Because there are $2m$ edge actions and $n$ node actions,
\begin{equation*}\begin{aligned}\left|W(\mathbf{x})-\bar W(\mathbf{x})\right|&=\frac{1}{2}\left|\left(Y_E(\mathbf{x})-\bar Y_E(\mathbf{x})\right)-\left(Y_P(\mathbf{x})-\bar Y_P(\mathbf{x})\right)\right|\\
\leq\frac{1}{2} \Big(\sum_{\{i,j\}\in E} & \sum_{r\in\{e_{ij},e_{ji}\}}\left|y_r^*(\mathbf{x})-\bar y_r(\mathbf{x})\right|+\sum_{i=1}^n\left|y_{p_i}^*(\mathbf{x})-\bar y_{p_i}(\mathbf{x})\right|\Big)\\
&\leq\frac{1}{2}\left(\frac{2m}{M}+\frac{n}{M}\right)=\frac{2m+n}{2M}.\end{aligned}\end{equation*}
\end{proof}

\begin{proof}[\textbf{Proof of Proposition~\ref{prop:sanitycheck}}]
    We verify Assumption~\ref{ass:active-tie-independence} before transferring the cubic maximum-cut separation. Fix any commitment $\mathbf{x}$ and let $\mathbf{h}(\mathbf{x})\in\mathbb{R}^K$ (recall $K = 2m + n + M$ is the number of follower pure actions and also the number of data samples) have coordinate $h_s(\mathbf{x})=1/2-d_s(\mathbf{x})$. From Equation~\eqref{eq:additive-hardness-follower-payoffs}, the follower sample-payoff vector generated by pure follower action $r$ is
\begin{equation}\label{eq:additive-hardness-pure-follower-vector}\mathbf{g}_F(\mathbf{x},r)=\mathbf{h}(\mathbf{x})+a\mathbf{e}_r,\end{equation}
where $\mathbf{e}_r$ is the standard basis vector indexed by sample $r$. The entropic penalty is finite for every $\mathbf{q}\in\Delta_K$. Recall from Equation~\eqref{eq:active-follower-actions} that $J_F(\mathbf{x},\mathbf{q})$ is the set of pure follower actions tied for the largest $\mathbf{q}$-weighted sample payoff. For the present construction,
\begin{equation}\label{eq:additive-hardness-active-actions}\mathbf{q}^{\top}\mathbf{g}_F(\mathbf{x},r)=\mathbf{q}^{\top}\mathbf{h}(\mathbf{x})+aq_r,\qquad J_F(\mathbf{x},\mathbf{q})=\argmax_{r\in A_F}q_r.\end{equation}
Suppose coefficients $c_r$, indexed by $J_F(\mathbf{x},\mathbf{q})$, satisfy $\sum_rc_r=0$ and $\sum_rc_r\mathbf{g}_F(\mathbf{x},r)=\mathbf{0}$. Substitution of Equation~\eqref{eq:additive-hardness-pure-follower-vector} make the first term $\mathbf{h}(\mathbf{x}) \sum_r c_r = 0$ and leaves $a\sum_rc_r\mathbf{e}_r=\mathbf{0}$. Linear independence of the standard basis vectors gives $c_r=0$ for every $r$. 
{\color{black}Thus, affine independence holds for every $\mathbf{q}\in\Delta_K$, and in particular for a dual optimizer of Problem~\eqref{eq:follower-minimax-dual} at each $\mathbf{x}$, proving Assumption~\ref{ass:active-tie-independence}.} The construction also uses entropic utilities, which satisfy Definition~\ref{def:convex-monetary-utility} and strict concavity modulo translations by Lemma~\ref{lem:entropic-properties}. 
\end{proof}

\begin{proof}[\textbf{Proof of Proposition~\ref{prop:entropy-follower-approximation}}]
Fix $\mathbf{x}$. The objective in Equation~\eqref{eq:entropy-regularized-follower-response} is continuous on the compact follower probability simplex, so a maximizer exists. The function $U_F(\mathbf{x},\cdot)$ is concave, while $\mathcal{H}$ is strictly concave on $\Delta(A_F)$. Their sum is therefore strictly concave for every $\tau>0$ and cannot have two distinct maximizers. This proves uniqueness.

Let $\mathbf{y}^*=\mathbf{y}^*(\mathbf{x})$ denote the unique unregularized response. Optimality of $\mathbf{y}_{\tau}=\mathbf{y}_{\tau}(\mathbf{x})$ for the regularized problem gives
\begin{equation}\label{eq:entropy-regularized-optimality-comparison}U_F(\mathbf{x},\mathbf{y}_{\tau})+\tau\mathcal{H}(\mathbf{y}_{\tau})\geq U_F(\mathbf{x},\mathbf{y}^*)+\tau\mathcal{H}(\mathbf{y}^*).\end{equation}
Entropy is nonnegative on the simplex and is at most $\log n_F$, with equality at the uniform strategy. Rearranging \eqref{eq:entropy-regularized-optimality-comparison} therefore yields
\begin{equation}\label{eq:entropy-follower-gap-upper-step}U_F(\mathbf{x},\mathbf{y}^*)-U_F(\mathbf{x},\mathbf{y}_{\tau})\leq\tau\bigl(\mathcal{H}(\mathbf{y}_{\tau})-\mathcal{H}(\mathbf{y}^*)\bigr)\leq\tau\log n_F.\end{equation}
The left side is nonnegative because $\mathbf{y}^*$ maximizes the original follower utility. This proves \eqref{eq:entropy-follower-gap}.
\end{proof}

\begin{proof}[Proof of Proposition~\ref{prop:implicit-gradient}]
For fixed $\tau>0$, define the KKT residual function
\[
\mathcal{R}_{\tau}(\mathbf{x},\mathbf{y},\kappa)=
\begin{bmatrix}
\nabla_{\mathbf{y}}U_F(\mathbf{x},\mathbf{y})-\tau(\mathbf{1}+\log\mathbf{y})-\kappa\mathbf{1}\\
\mathbf{1}^{\top}\mathbf{y}-1
\end{bmatrix}.
\]
The first-order conditions in Equation~\eqref{eq:entropy-regularized-follower-kkt} are equivalent to $\mathcal{R}_{\tau}(\mathbf{x},\mathbf{y}_{\tau}(\mathbf{x}),\kappa_{\tau}(\mathbf{x}))=\mathbf{0}$. The Jacobian of $\mathcal{R}_{\tau}$ with respect to $(\mathbf{y},\kappa)$, evaluated at $(\mathbf{x},\mathbf{y}_{\tau}(\mathbf{x}),\kappa_{\tau}(\mathbf{x}))$, is
\[
D_{(\mathbf{y},\kappa)}\mathcal{R}_{\tau}
=
\begin{bmatrix}
\mathbf{B}_{\tau}(\mathbf{x})&-\mathbf{1}\\
\mathbf{1}^{\top}&0
\end{bmatrix}.
\]
The upper-left block is $\mathbf{B}_{\tau}(\mathbf{x})$ because differentiating $-\tau\log\mathbf{y}$ gives $-\tau\operatorname{Diag}(\mathbf{y})^{-1}$.

We next prove that this Jacobian is nonsingular. Suppose that $(\mathbf{v},c)$ belongs to its null space. Then $\mathbf{B}_{\tau}(\mathbf{x})\mathbf{v}-c\mathbf{1}=\mathbf{0}$ and $\mathbf{1}^{\top}\mathbf{v}=0$. Multiplying the first equality by $\mathbf{v}^{\top}$ gives $\mathbf{v}^{\top}\mathbf{B}_{\tau}(\mathbf{x})\mathbf{v}=0$. If $\mathbf{v}\neq\mathbf{0}$, however,
\[
\mathbf{v}^{\top}\mathbf{B}_{\tau}(\mathbf{x})\mathbf{v}
=
\mathbf{v}^{\top}\nabla^2_{\mathbf{y}\mathbf{y}}U_F(\mathbf{x},\mathbf{y}_{\tau})\mathbf{v}
-\tau\sum_{j\in A_F}\frac{v_j^2}{y_{\tau,j}(\mathbf{x})}<0.
\]
The first term is nonpositive by concavity of $U_F(\mathbf{x},\cdot)$, while the second is strictly negative because $\tau>0$ and every coordinate of $\mathbf{y}_{\tau}(\mathbf{x})$ is positive. Hence $\mathbf{v}=\mathbf{0}$, and then $c=0$. Thus the Jacobian is nonsingular.

The implicit-function theorem therefore makes $(\mathbf{y}_{\tau}(\mathbf{x}),\kappa_{\tau}(\mathbf{x}))$ locally continuously differentiable. We may now differentiate the identity $\mathcal{R}_{\tau}(\mathbf{x},\mathbf{y}_{\tau}(\mathbf{x}),\kappa_{\tau}(\mathbf{x}))=\mathbf{0}$ with respect to $\mathbf{x}$. The chain rule gives
\[
\begin{bmatrix}
\mathbf{B}_{\tau}(\mathbf{x})&-\mathbf{1}\\
\mathbf{1}^{\top}&0
\end{bmatrix}
\begin{bmatrix}
D_{\mathbf{x}}\mathbf{y}_{\tau}(\mathbf{x})\\
D_{\mathbf{x}}\kappa_{\tau}(\mathbf{x})
\end{bmatrix}
+
\begin{bmatrix}
\nabla^2_{\mathbf{y}\mathbf{x}}U_F(\mathbf{x},\mathbf{y}_{\tau}(\mathbf{x}))\\
\mathbf{0}^{\top}
\end{bmatrix}
=\mathbf{0},
\]
which is equivalent to Equation~\eqref{eq:implicit-response-linear-system}. Uniqueness of the regularized response and its multiplier makes the local functions agree wherever their domains overlap. Finally, applying the chain rule to $\Phi_{\tau}(\mathbf{x})=U_L\!\left(\mathbf{x},\mathbf{y}_{\tau}(\mathbf{x})\right)$ gives Equation~\eqref{eq:smoothed-leader-implicit-gradient}.
\end{proof}

\begin{proof}[\textbf{Proof of Theorem~\ref{thm:implicit-gradient-approximation}}]
The regularized response map $\mathbf{x}\mapsto\mathbf{y}_{\tau}(\mathbf{x})$ is continuous by Proposition~\ref{prop:implicit-gradient}, while the original response map $\mathbf{x}\mapsto\mathbf{y}^*(\mathbf{x})$ is continuous by Theorem~\ref{thm:convex-risk-attainment}. Compactness of the leader probability simplex therefore ensures that the displayed maximum defining $d_{\tau}$ exists.

We first show that, for every $\mathbf{x}$ and $\mathbf{y},\mathbf{y}'\in\Delta(A_F)$, the leader's utility satisfies $|U_L(\mathbf{x},\mathbf{y})-U_L(\mathbf{x},\mathbf{y}')|\leq\|\mathbf{y}-\mathbf{y}'\|_1$. We will apply this bound to $\mathbf{y}_{\tau}(\mathbf{x})$ and $\mathbf{y}^*(\mathbf{x})$. Let $X$ and $Y$ be two random rewards and set $a=\lVert X-Y\rVert_{\infty}$. Then $X\leq Y+a$ and $Y\leq X+a$. Monotonicity and translation equivariance imply $\rho_L(X)\leq\rho_L(Y)+a$ and $\rho_L(Y)\leq\rho_L(X)+a$, respectively. Hence $|\rho_L(X)-\rho_L(Y)|\leq\lVert X-Y\rVert_{\infty}$. For fixed $\mathbf{x}$ and $k$, every coefficient $\sum_i x_i u_L(i,j\mid\xi_k)$ lies in $[0,1]$. The mixed-payoff formula therefore gives
\begin{equation}\label{eq:leader-payoff-response-lipschitz}\left|g_{L,k}(\mathbf{x},\mathbf{y})-g_{L,k}(\mathbf{x},\mathbf{y}')\right|\leq\sum_{j\in A_F}|y_j-y_j'|=\lVert\mathbf{y}-\mathbf{y}'\rVert_1.\end{equation}
Applying $|\rho_L(X)-\rho_L(Y)|\leq\|X-Y\|_{\infty}$ to the random rewards induced by $(\mathbf{x},\mathbf{y})$ and $(\mathbf{x},\mathbf{y}')$, and using Equation~\eqref{eq:leader-payoff-response-lipschitz}, gives
\begin{equation}\label{eq:leader-utility-response-lipschitz}\left|U_L(\mathbf{x},\mathbf{y})-U_L(\mathbf{x},\mathbf{y}')\right|\leq\lVert\mathbf{y}-\mathbf{y}'\rVert_1.\end{equation}

Let $\mathbf{x}^*$ attain $V$. Because Algorithm~\ref{alg:implicit-gradient} selects the returned commitment having the largest original-game leader value, Equation~\eqref{eq:leader-utility-response-lipschitz} and the global optimality of $\mathbf{x}^{g}$ for $\Phi_{\tau}$ give
\begin{align*}
U_L(\widehat{\mathbf{x}},\widehat{\mathbf{y}})
&\geq U_L\!\big(\mathbf{x}^{g},\mathbf{y}^*(\mathbf{x}^{g})\big)\\
&\geq \Phi_{\tau}(\mathbf{x}^{g})-d_{\tau}\\
&\geq \Phi_{\tau}(\mathbf{x}^*)-d_{\tau}\\
&\geq U_L\!\left(\mathbf{x}^*,\mathbf{y}^*(\mathbf{x}^*)\right)-2d_{\tau}
=V-2d_{\tau}.
\end{align*}
This proves Equation~\eqref{eq:implicit-gradient-general-guarantee}.

We next prove that $d_{\tau}\to0$ as $\tau \to 0$. Since $0\leq\mathcal{H}(\mathbf{y})\leq\log n_F$, the regularized follower objective converges uniformly to $U_F$ as $\tau\to 0$. Suppose, to the contrary, that there exist $a>0$, $\tau_n \to 0$, and $\mathbf{x}^n$ such that $\|\mathbf{y}_{\tau_n}(\mathbf{x}^n)-\mathbf{y}^*(\mathbf{x}^n)\|_1\geq a$. By compactness, along a subsequence, $\mathbf{x}^n\to\overline{\mathbf{x}}$ and $\mathbf{y}_{\tau_n}(\mathbf{x}^r)\to\overline{\mathbf{y}}$. For every $\mathbf{y}\in\Delta(A_F)$, optimality of the regularized response gives
\begin{equation}
U_F(\mathbf{x}^n,\mathbf{y}_{\tau_r}(\mathbf{x}^n))+\tau_r\mathcal{H}(\mathbf{y}_{\tau_n}(\mathbf{x}^r))
\geq U_F(\mathbf{x}^n,\mathbf{y})+\tau_n\mathcal{H}(\mathbf{y}).
\end{equation}
Taking limits shows that $\overline{\mathbf{y}}\in BR_F(\overline{\mathbf{x}})$, so uniqueness gives $\overline{\mathbf{y}}=\mathbf{y}^*(\overline{\mathbf{x}})$. Continuity of $\mathbf{y}^*$ then implies $\mathbf{y}^*(\mathbf{x}^n)\to\overline{\mathbf{y}}$, contradicting the lower bound by $a$. Hence $d_{\tau}\to0$ as $\tau \to 0$.

Finally, suppose the follower utility is uniformly $\mu$-strongly concave. Strong concavity and the first-order optimality condition at $\mathbf{y}^*(\mathbf{x})$ imply, for every feasible $\mathbf{y}$,
\begin{equation*}\label{eq:strong-concavity-response-growth}U_F\!\left(\mathbf{x},\mathbf{y}^*(\mathbf{x})\right)-U_F(\mathbf{x},\mathbf{y})\geq\frac{\mu}{2}\left\lVert\mathbf{y}-\mathbf{y}^*(\mathbf{x})\right\rVert_2^2.\end{equation*}
Apply the above to $\mathbf{y}=\mathbf{y}_{\tau}(\mathbf{x})$ and combine it with approximation of follower response in Equation~\eqref{eq:entropy-follower-gap}. This yields
\begin{equation*}\left\lVert\mathbf{y}_{\tau}(\mathbf{x})-\mathbf{y}^*(\mathbf{x})\right\rVert_2\leq\sqrt{\frac{2\tau\log n_F}{\mu}}.\end{equation*}
The norm inequality $\lVert\mathbf{v}\rVert_1\leq\sqrt{n_F}\lVert\mathbf{v}\rVert_2$ then bounds $d_{\tau}$ by the square-root term in Equation~\eqref{eq:implicit-gradient-explicit-guarantee}. Substituting this bound into Equation~\eqref{eq:implicit-gradient-general-guarantee} proves Equation~\eqref{eq:implicit-gradient-explicit-guarantee}. The displayed choice of $\tau$ makes its error term at most $\epsilon$.
\end{proof}

\section{Additional Examples and Results}\label{app:additionalexample}

Another standard example is the \emph{utility-based shortfall utility}~\cite{foellmer2010convex}. For an increasing and strictly convex function $\varphi:\mathbb{R}\to\mathbb{R}$, define $\rho_\varphi(X)=\sup\{m\in\mathbb{R}:\sum_{k=1}^Kp_k\varphi(m-X(\xi_k))\leq\varphi(0)\}$. Thus, $\rho_\varphi(X)$ is the largest certain reward whose expected transformed shortfall remains acceptable. Its penalized-distribution representation is
\begin{equation}\begin{aligned}\rho_\varphi(X)=\min_{\mathbf{q}\in\Delta_K}\left\{\sum_{k=1}^Kq_kX(\xi_k)+\alpha_\varphi(\mathbf{q})\right\},\\
\qquad \alpha_\varphi(\mathbf{q})=\inf_{s>0}s\left\{\varphi(0)+\sum_{k=1}^Kp_k\varphi^*\left(\frac{q_k}{sp_k}\right)\right\},\end{aligned}\end{equation}
where $\varphi^*(v)=\sup_{t\in\mathbb{R}}\{vt-\varphi(t)\}$. For example, $\varphi(t)=\log(1+\exp(t))$ gives a smooth utility whose value is the unique solution of $\sum_{k=1}^Kp_k\varphi(m-X(\xi_k))=\varphi(0)$.

\begin{example}[A uniquely mixed entropic response]\label{ex:entropic-mixed-response}
Let $A_L=A_F=\{1,2\}$, $\mathbf{x}=(p,1-p)$, and $\mathbf{y}=(q,1-q)$. Consider $K=2$ equally likely samples and set $u_L(i,j\mid\xi_k)=\mathbf{1}_{\{i=k\}}$ and $u_F(i,j\mid\xi_k)=(1+\mathbf{1}_{\{j=k\}}-\mathbf{1}_{\{i=k\}})/2$. Both players use entropic utility with parameters $\theta_L,\theta_F>0$. Thus, for each $\nu\in\{L,F\}$, $U_\nu(\mathbf{x},\mathbf{y})=-\theta_\nu^{-1}\log((e^{-\theta_\nu g_{\nu,1}(\mathbf{x},\mathbf{y})}+e^{-\theta_\nu g_{\nu,2}(\mathbf{x},\mathbf{y})})/2)$. The follower's sample rewards are $g_{F,1}(\mathbf{x},\mathbf{y})=(1+q-p)/2$ and $g_{F,2}(\mathbf{x},\mathbf{y})=(1+p-q)/2$. Since $e^{-\theta(1+d)/2}+e^{-\theta(1-d)/2}\geq2e^{-\theta/2}$, with equality only at $d=0$, the unique follower response is $q=p$. In particular, at $p=1/2$, the mixed response $q=1/2$ gives the constant reward $1/2$ and is strictly preferred to the pure responses, whose sample rewards are $(3/4,1/4)$ and $(1/4,3/4)$. The leader's sample rewards are $g_{L,1}(\mathbf{x},\mathbf{y})=p$ and $g_{L,2}(\mathbf{x},\mathbf{y})=1-p$, so the same inequality gives the unique optimistic equilibrium $(p^*,q^*)=(1/2,1/2)$. 
\end{example}

\begin{example}[A noninducible optimistic CVaR response]\label{ex:cvar-noninducible-response}
Let $A_L=\{L_1,L_2\}$ and $A_F=\{a,b\}$, with $\mathbf{x}=(p,1-p)$ and $\mathbf{y}=(q,1-q)$. The mean--CVaR utility of a random reward $X$ is $\rho_{\mathrm{CVaR}}(X)=(1-\gamma_c)\mathbb{E}[X]+\gamma_c\operatorname{CVaR}_{\alpha}(X)$, where $\gamma_c\in[0,1]$ and
\begin{equation}\label{eq:cvar-inducibility-utility}
\operatorname{CVaR}_{\alpha}(X)=\sup_{z\in\mathbb{R}}\left\{z+\frac{1}{\alpha}\mathbb{E}\!\left[\min\{0,X-z\}\right]\right\}.
\end{equation}
There are two equally likely samples, and both players use $\gamma_c=1$ and $\alpha=1/2$. Thus, for two equally likely realizations $z_1,z_2$, their utility is $\operatorname{CVaR}_{1/2}(z_1,z_2)=\min\{z_1,z_2\}$. The follower payoff matrices are
\begin{equation*}
\bigl(u_F(i,j\mid\xi_1)\bigr)_{ij}=\begin{pmatrix}7/8&1/4\\5/8&1/2\end{pmatrix},\qquad \bigl(u_F(i,j\mid\xi_2)\bigr)_{ij}=\begin{pmatrix}3/8&1/4\\1/8&1/2\end{pmatrix}.
\end{equation*}
The resulting sample payoffs are $g_{F,1}(p,q)=1/2-p/4+q(1/8+p/2)$ and $g_{F,2}(p,q)=1/2-p/4+q(-3/8+p/2)$. Since $g_{F,1}(p,q)-g_{F,2}(p,q)=q/2\geq0$, the follower utility is $U_F(p,q)=g_{F,2}(p,q)=1/2-p/4+q(p/2-3/8)$. Therefore, $BR_F(p)=\{0\}$ for $p<3/4$, $BR_F(3/4)=[0,1]$, and $BR_F(p)=\{1\}$ for $p>3/4$.

Let the leader payoff matrices at $\xi_1$ and $\xi_2$ be $\left(\begin{smallmatrix}1&0\\1&0\end{smallmatrix}\right)$ and $\left(\begin{smallmatrix}0&1\\0&1\end{smallmatrix}\right)$, respectively. The leader's sample payoffs are $q$ and $1-q$, so $U_L(q)=\min\{q,1-q\}=1/2-|q-1/2|$. Consequently, the unique optimistic Stackelberg pair is $(p^*,q^*)=(3/4,1/2)$. It is not inducible because every unique follower response at a commitment different from $p^*$ belongs to $\{0,1\}$, so no sequence of such responses converges to $q^*=1/2$. At $p^*$, actions $a$ and $b$ induce the distinct random rewards taking values $(13/16,5/16)$ and $(5/16,5/16)$, respectively, yet both have mean--CVaR utility equal to $5/16$.
\end{example}

The following results shows how positive homogeneity destroys strict concavity:
\begin{proposition}[Positive homogeneity precludes strict curvature]\label{prop:coherence-precludes-strict-curvature}
No coherent utility is strictly concave modulo translations on any convex domain containing two distinct positive multiples of the same nonconstant random reward.
\end{proposition}

\begin{proof}
Let $X$ be nonconstant and suppose the domain contains $X$ and $tX$ for some $t>0$ with $t\neq1$. Their difference $(1-t)X$ is not almost surely constant. For any $\lambda\in(0,1)$, positive homogeneity gives
\begin{equation*}\label{eq:coherent-ray-equality}\rho\!\left(\lambda X+(1-\lambda)tX\right)=\bigl(\lambda+(1-\lambda)t\bigr)\rho(X)=\lambda\rho(X)+(1-\lambda)\rho(tX).\end{equation*}
The strict inequality in Equation~\eqref{eq:strict-concavity-modulo-translations} therefore fails. 
\end{proof}

\section{Experimental Setting Details}\label{app:risk-aware-guards}

AI was used for coding, so we provide the disclosure here. 
\paragraph{Disclosure of AI-assisted Technologies.}
OpenAI GPT and Codex (GPT-5.6) was used to assist with literature discovery, checking mathematical derivations, implementing the algorithms and experiments, debugging newly written data-processing and plotting code, and minor editing of the exposition. The third-party GUARD source code of \citet{krever2025guard} was obtained from its authors and was neither generated nor modified by Codex. The authors determined the research questions, model, assumptions, final mathematical claims, and experimental protocol; checked the cited sources and mathematical arguments; executed and validated the code; and independently verified the reported results. The authors take full responsibility for all content. Some example prompts are ``Implement the wildlife-security-game experiment as discussed using the downloaded GUARD data and model. Preserve the original schedules and payoff construction, construct environmental payoff samples, store all experimental results, and keep plotting code separate from optimization code,'' ``Review the submission PDF for mathematical inconsistencies, unsupported claims, citation errors, experimental discrepancies, typographical errors, and anonymity issues,'' and ``Code the synthetic scalability experiment as discussed comparing the implicit-gradient method with Gurobi across varying numbers of follower actions and environmental samples. Use fixed random seeds, five instances, ten gradient restarts, and retain sufficient data for runtime and solution-quality comparisons.''

\paragraph{\textbf{Scalability Experiments Detailed Setting}} First, for the Gurobi implementation, we replace each complementarity constraint in Equation~\eqref{eq:convex-risk-mpec} by the exact big-$M$ reformulation $0\leq y_j\leq Mz_j$ and $0\leq\kappa_F-\sum_{k=1}^Kq_{F,k}\sum_{i\in A_L}x_i u_F(i,j\mid\xi_k)\leq M(1-z_j)$, where $z_j\in\{0,1\}$; the payoff normalization makes $M=1$ valid for both inequalities.

We implement a canonical Stackelberg security game~\cite{tambe2011}. Recall we have $n_F\in\{40, 100, 200, 400\}$ targets and $K\in\{50,100,200\}$, with $r = n_F/2$ defender resources. For each target $t$, we draw a base defender loss $d_t\sim\operatorname{Unif}(0.35,0.90)$ and attacker gain $a_t\sim\operatorname{Unif}(0.30,0.95)$. In environmental sample $k$, these values become $d_{kt}=\operatorname{clip}(d_t+\zeta_k+\epsilon^L_{kt},0.05,0.98)$ and $a_{kt}=\operatorname{clip}(a_t+0.65\zeta_k+\epsilon^F_{kt},0.02,0.98)$, where $\zeta_k\sim\mathcal N(0,0.08^2)$ is shared across targets and players, while $\epsilon^L_{kt}\sim\mathcal N(0,0.06^2)$ and $\epsilon^F_{kt}\sim\mathcal N(0,0.07^2)$ are independent. We independently draw the effectiveness of covering target $t$ as $\eta_{kt}=0.60+0.38B_{kt}$, where $B_{kt}\sim\operatorname{Beta}(6,2.5)$. The attacker always selects one target. If its attack on target $t$ succeeds, the defender and attacker receive $1-d_{kt}$ and $a_{kt}$; if coverage defeats the attack, they receive $1$ and $0$. Like in standard Stackelberg games, we use marginal coverage probabilities, which are the defender's variable of optimization. Consequently, under marginal coverage probability $x_t$, their sample rewards are $1-d_{kt}(1-x_t\eta_{kt})$ and $a_{kt}(1-x_t\eta_{kt})$, respectively. As is standard, $0\leq x_t\leq1,\; \sum_t x_t\leq r.$ All $K$ environmental samples receive probability $1/K$. We use $\zeta=10^{-4}$ and $c=10^{-4}$. The schedule of $\tau$ and iteration counts is shown in Table~\ref{tab:tauschedule}.

Note that in the scalability experiments, $\mathbf{x}$ denotes the marginal target-coverage vector rather than a probability distribution over pure leader actions, and therefore satisfies $0\leq x_t\leq1$ and $\sum_{t=1}^{n_F}x_t\leq r$; this set is the convex hull of allocations covering at most $r$ targets, so Algorithm~\ref{alg:implicit-gradient} projects onto this compact coverage polytope, while its follower regularization and implicit-derivative calculation remain unchanged.

{\color{black}Assumption~\ref{ass:active-tie-independence} is not imposed on the synthetic instance generator. Thus, these benchmarks evaluate the computational methods beyond the sufficient conditions of Theorem~\ref{thm:convex-risk-attainment}: the general MPEC and the positive-entropy implicit derivative remain valid without that assumption. The original-game approximation guarantee in Theorem~\ref{thm:implicit-gradient-approximation} retains its stated theoretical conditions.}
\begin{table}[t]
\centering
\caption{Continuation schedule used in the experiments.} \label{tab:tauschedule}
\label{tab:continuation-schedule}
\begin{tabular}{c@{\qquad}c@{\qquad}r}
\toprule
Stage $s$ & $\tau_s$ & $T_s$\\
\midrule
1 & $10^{-3}$ & 7500\\
2 & $3\times10^{-4}$ & 500\\
3 & $10^{-4}$ & 500\\
4 & $3\times10^{-5}$ & 500\\
5 & $10^{-5}$ & 500\\
6 & $3\times10^{-6}$ & 500\\
7 & $10^{-6}$ & 10000\\
\bottomrule
\end{tabular}
\end{table}

\paragraph{\textbf{Real Data Experiment Detailed Setting}} This part gives the complete construction of the real-world security game used in our experiments. We first describe the  GUARD framework of \citet{krever2025guard}, and then decompose its target scores into environmental payoff samples for our \textsc{Risk-Aware-Guards} model. We use the GUARD source and schedule-generation procedure without modification, subsequently apply the exact action-set reductions described below, and introduce environmental uncertainty only through the payoff samples.

\paragraph{The GUARD Green Security Game.} We first describe the prior work GUARD~\cite{krever2025guard}. GUARD constructs a security game from animal-location observations and a spatial grid. The defender controls mobile patrol resources, while the attacker selects a target containing valuable wildlife. We identify the defender with the Stackelberg leader and the attacker with the follower. The experiment concerns Lob\'ek\'e National Park in southeastern Cameroon and uses GPS-collar observations from the \emph{Elephant Research--Lob\'ek\'e National Park (Cameroon)} study, obtained through the Movebank animal-movement repository. The Lob\'ek\'e instance uses a $7\times7$ grid, ten target cells obtained by centroid-based clustering, two defender resources, and one attacker. Each defender may begin from any of three specified patrol bases. A patrol must return to its base within seven movement steps and must spend one step at a target to cover it. In the simple schedule, a schedule covers one target and has zero movement cost. In the general schedule, a schedule may cover several feasible targets and uses GUARD's step-cost parameter $1.17$. GUARD generates 13 base-specific schedule choices per defender in the simple schedule and 52 in the general schedule. Because a pure defender action chooses one schedule for each resource, the expanded leader action sets have $13^2=169$ and $52^2=2704$ actions, respectively. The follower has one pure action for each of the ten targets.

Let $T=\{1,\ldots,10\}=A_F$ be the target set, and let $A_L$ be the expanded set of joint defender schedules. For $i\in A_L$ and $j\in T$, let $C_{ij}\in\{0,1\}$ indicate whether at least one defender schedule in joint action $i$ covers target $j$. Let $\kappa_i\geq0$ be the joint patrol cost returned by GUARD; $\kappa_i=0$ in the simple schedule. In the general schedule, GUARD multiplies the input step-cost parameter $1.17$ by the mean geographic side length of a grid cell, multiplies the result by the number of movement steps in each schedule, and sums the two schedule costs to obtain $\kappa_i$.

GUARD first removes observations with missing fields and then retains coordinates in the latitude interval $[2.0530,2.2837]$ and longitude interval $[15.8790,16.2038]$. The combined file has 3,182 complete observations from five tracked animals; 1,592 observations from four animals remain inside this Lobeke rectangle. K-means with ten clusters is applied to these retained coordinates. If $N_j$ is the number of retained observations assigned to cluster $j$, $N_{\mathrm{obs}}=1592$, and $N_{\mathrm{ani}}=4$, GUARD assigns target $j$ the score
\begin{equation}\label{eq:guard-aggregate-score}\bar s_j=\frac{N_{\mathrm{ani}}N_j}{N_{\mathrm{obs}}}=\frac{4N_j}{1592}.
\end{equation}
In target order, the cluster counts are $286$, $90$, $146$, $208$, $97$, $105$, $235$, $241$, $48$, and $136$. Thus $\sum_{j=1}^{10}\bar s_j=4$. The value four is the number of tracked animals represented after geographic filtering, not the animal population of the park.

The attacker's value also depends on proximity to the designated escape boundary, which represents a route through which the attacker can leave the protected area. In the Lob\'ek\'e instance, GUARD represents the nearby Sangha River as this boundary. Let $d_j$ be GUARD's distance from target $j$ to that boundary, and let $d_{\min}$ and $d_{\max}$ be the minimum and maximum of these distances over the ten targets. GUARD uses the multiplier
\begin{equation}\label{eq:guard-escape-multiplier}\beta_j=1+\alpha\left(1-\frac{d_j-d_{\min}}{d_{\max}-d_{\min}}\right).
\end{equation}
This multiplier makes a target more attractive to the attacker when it offers an easier escape. With GUARD's experimental choice $\alpha=1$, the closest target has $\beta_j=2$, the farthest has $\beta_j=1$, and the remaining targets receive intermediate values. Escape proximity affects only the attacker's valuation; the defender's loss depends on the wildlife value at the target.

GUARD converts the target score into monetary value using two domain-calibrated constants. The attacker value $2350$ approximates the black-market value in U.S. dollars of two elephant tusks, whereas the defender value $22966$ represents the estimated one-year ecotourism value of an elephant. Consequently, the aggregate uncovered target values are
\begin{equation}\label{eq:guard-aggregate-target-values}\bar v_{Fj}=2350\,\beta_j\bar s_j,\qquad \bar v_{Lj}=-22966\,\bar s_j.
\end{equation}
The first expression is positive because a successful attack benefits the attacker. The second is negative because an uncovered attack imposes the corresponding wildlife loss on the defender.
Coverage divides the magnitude of each player's target value by five. Define $h_{ij}=1-4C_{ij}/5$, so $h_{ij}=1$ when target $j$ is uncovered by defender action $i$ and $h_{ij}=1/5$ when it is covered. GUARD then constructs its deterministic payoff matrices in three steps.
\begin{enumerate}
\item \emph{Normalize the target values.} GUARD divides every attacker and defender target value by the same constant
\begin{equation}\label{eq:guard-first-normalization}B=\max\left\{\max_{j\in T}|\bar v_{Fj}|,\max_{j\in T}|\bar v_{Lj}|\right\}=16503.206030150755.
\end{equation}
This preserves the signs and relative magnitudes of the target values while making their largest absolute value equal to one.
\item \emph{Apply coverage and patrol cost.} For each defender schedule $i$ and attacked target $j$, GUARD forms
\begin{equation}\label{eq:guard-prefinal-payoffs}\bar w_F(i,j)=\frac{h_{ij}\bar v_{Fj}}{B},\qquad \bar w_L(i,j)=\frac{h_{ij}\bar v_{Lj}}{B}-\kappa_i.
\end{equation}
Thus coverage scales both target payoffs by $h_{ij}$, while the defender additionally pays the patrol cost $\kappa_i$.
\item \emph{Normalize the complete game.} After coverage and patrol costs have been incorporated, GUARD computes
\begin{equation}\label{eq:guard-final-normalization}H=\max\left\{\max_{i\in A_L,\,j\in A_F}|\bar w_F(i,j)|,\max_{i\in A_L,\,j\in A_F}|\bar w_L(i,j)|\right\}
\end{equation}
and returns the follower and leader payoff matrices
\begin{equation}\label{eq:guard-final-payoffs}\bar u_F(i,j)=\frac{\bar w_F(i,j)}{H},\qquad \bar u_L(i,j)=\frac{\bar w_L(i,j)}{H}.
\end{equation}
This second division places both matrices on one common scale with maximum absolute entry one; it does not shift either matrix into $[0,1]$.
\end{enumerate}
We use the simple and general patrol schedules generated by GUARD’s original code under the fixed parameters described above. In the simple schedule, every patrol cost is zero. Because the first normalization already bounds all target-dependent entries by one and an entry of magnitude one is attained, $H=1$. In the general schedule, schedules can visit multiple targets and incur positive movement costs. These costs are subtracted after the target-value normalization and dominate the target-dependent terms for the longest schedules, giving $H=83.4259918063$. The values of $H$ therefore differ because the two schedules have different patrol costs and schedule sets, not because their wildlife data or target values differ.

\paragraph{Environmental Samples.} For our risk-aware setting, we extract data samples from the original GUARDS data. The deterministic GUARD score $\bar s_j$ aggregates all retained locations. To expose the underlying environmental uncertainty without changing targets or schedules, we let $\xi_k$, $k=1,\ldots,10$, denote the state in which a representative retained animal-location observation belongs to target cluster $k$. Its nominal probability is the empirical cluster frequency
\begin{equation}\label{eq:risk-aware-guards-state-probability}p_k=\frac{N_k}{N_{\mathrm{obs}}}=\frac{N_k}{1592}>0,\qquad \sum_{k=1}^{10}p_k=1.
\end{equation}
The probabilities are nonuniform because the cluster counts in \eqref{eq:guard-aggregate-score} differ. All observations assigned to the same cluster produce the same payoff matrix in the construction below, so combining them into one state preserves their total probability.

In state $\xi_k$, define the target-score realization
\begin{equation}\label{eq:risk-aware-guards-sample-score}s_j(\xi_k)=N_{\mathrm{ani}}\mathbf{1}_{\{j=k\}}=4\mathbf{1}_{\{j=k\}}.
\end{equation}
This value is an importance-scaled observation; it does not assert that four animals simultaneously occupy target $k$. Its purpose is to retain GUARD's scale while decomposing the aggregate score, since
\begin{equation}\label{eq:risk-aware-guards-score-reconstruction}\sum_{k=1}^{10}p_k s_j(\xi_k)=4p_j=\frac{4N_j}{1592}=\bar s_j.
\end{equation}
Equivalently, one could assign score one to the active target and multiply both animal-value constants in \eqref{eq:guard-aggregate-target-values} by four; \eqref{eq:risk-aware-guards-sample-score} instead leaves GUARD's constants unchanged.

For every state, define $v_{Fj}(\xi_k)=2350\,\beta_j s_j(\xi_k)$ and $v_{Lj}(\xi_k)=-22966\,s_j(\xi_k)$. We keep the aggregate normalization constants $B$ and $H$, the coverage indicators $C_{ij}$, and the action costs $\kappa_i$ fixed across states. In particular, we do not renormalize any environmental state separately. The signed GUARD-scale sample payoffs are
\begin{equation}\label{eq:risk-aware-guards-signed-payoffs}r_F(i,j\mid\xi_k)=\frac{h_{ij}v_{Fj}(\xi_k)}{BH},\qquad r_L(i,j\mid\xi_k)=\frac{h_{ij}v_{Lj}(\xi_k)/B-\kappa_i}{H}.
\end{equation}
Equations~\eqref{eq:risk-aware-guards-score-reconstruction}--\eqref{eq:risk-aware-guards-signed-payoffs} imply the exact reconstruction identities
\begin{equation}\label{eq:risk-aware-guards-payoff-reconstruction}\begin{aligned}\sum_{k=1}^{10}p_k r_F(i,j\mid\xi_k)&=\bar u_F(i,j),\\ \sum_{k=1}^{10}p_k r_L(i,j\mid\xi_k)&=\bar u_L(i,j),\qquad i\in A_L,\ j\in A_F.\end{aligned}\end{equation}
For the leader, the identity uses $\sum_kp_k=1$ because the patrol cost is incurred in every environmental state. Thus the original deterministic GUARD game is precisely the risk-neutral expectation of the constructed signed samples.

\paragraph{Normalized \textsc{Risk-Aware-Guards} game.} The theoretical model assumes rewards in $[0,1]$. For each schedule configuration, let $\underline r_F=\min_{k,i,j}r_F(i,j\mid\xi_k)$, $\overline r_F=\max_{k,i,j}r_F(i,j\mid\xi_k)$, $\underline r_L=\min_{k,i,j}r_L(i,j\mid\xi_k)$, and $\overline r_L=\max_{k,i,j}r_L(i,j\mid\xi_k)$. We apply one state-independent affine map to each player's complete sample-payoff tensor:
\begin{equation}\label{eq:risk-aware-guards-normalized-payoffs}\begin{aligned}u_F(i,j\mid\xi_k)&=\frac{r_F(i,j\mid\xi_k)-\underline r_F}{\overline r_F-\underline r_F},\\ u_L(i,j\mid\xi_k)&=\frac{r_L(i,j\mid\xi_k)-\underline r_L}{\overline r_L-\underline r_L}.\end{aligned}\end{equation}
Both $u_F(i,j\mid\xi_k)$ and $u_L(i,j\mid\xi_k)$ therefore lie in $[0,1]$.
The normalization bounds are
\begin{equation}\label{eq:risk-aware-guards-normalization-bounds}\begin{aligned}
\text{simple leader: }&(\underline r_L,\overline r_L)=(-5.5664335664,0),\\
\text{simple follower: }&(\underline r_F,\overline r_F)=(0,1.1391725926),\\
\text{general leader: }&(\underline r_L,\overline r_L)=(-1.0547363414,-0.2470033320),\\
\text{general follower: }&(\underline r_F,\overline r_F)=(0,0.0136548882).
\end{aligned}\end{equation}
Because the map in \eqref{eq:risk-aware-guards-normalized-payoffs} is fixed across states, the expected normalized matrix is the same affine transformation of the stored deterministic GUARD matrix in \eqref{eq:risk-aware-guards-payoff-reconstruction}; state-by-state normalization would not have this property and would change the risk-sensitive game.

The resulting \textsc{Risk-Aware-Guards} instances use the probabilities in \eqref{eq:risk-aware-guards-state-probability} and the payoff samples in \eqref{eq:risk-aware-guards-normalized-payoffs} within the general model of Section~\ref{sec:convex-preliminaries}. The simple and general payoff tensors have dimensions $10\times169\times10$ and $10\times2704\times10$, respectively, with indices corresponding to environmental state, leader action, and follower action. For mixed strategies $(\mathbf{x},\mathbf{y})$, the sample reward is exactly Equation~\eqref{eq:convex-sample-payoff}, and each player evaluates the resulting scalar-valued random reward using its own convex-risk utility. In particular, our entropic experiments use \eqref{eq:entropic-utility} with the nonuniform probabilities $p_k$ above and player-specific parameters $\theta_L$ and $\theta_F$, both interpreted on the normalized payoff scale in \eqref{eq:risk-aware-guards-normalized-payoffs}.

{\color{black} The payoff construction also satisfies Assumption~\ref{ass:active-tie-independence}. Since $\underline r_F=0$, Equations~\eqref{eq:risk-aware-guards-sample-score}, \eqref{eq:risk-aware-guards-signed-payoffs}, and \eqref{eq:risk-aware-guards-normalized-payoffs} give
\begin{equation*}g_{F,k}(\mathbf{x},\mathbf{e}_j)=\frac{9400\beta_j}{BH\overline r_F}\left(\sum_{i\in A_L}x_i h_{ij}\right)\mathbf{1}_{\{k=j\}}.\end{equation*}
Here $h_{ij}\geq1/5$, so $\sum_i x_i h_{ij}\geq1/5$. Every pure follower action therefore induces a positive multiple of a different coordinate vector. A linear combination of these vectors can vanish only if every coefficient is zero, proving linear independence and hence affine independence of every tied subset, including one selected by a dual optimizer. This holds for every leader strategy in both schedule configurations and is preserved by the leader-action reduction below. For $\theta_F>0$, entropic utility also satisfies strict concavity modulo translations, so Theorem~\ref{thm:convex-risk-attainment} applies. 
}

\paragraph{Exact Leader-Action Reduction.} GUARD distinguishes joint patrol actions by the schedules assigned to the two labeled defenders, but its payoffs retain only their combined coverage and total patrol cost. Let $\mathbf{C}_i=(C_{i1},\ldots,C_{i,10})$ be the coverage vector of action $i$, and recall that $\kappa_i$ is its patrol cost. For example, the simple instance contains the two assignments $(\{0\},\{3\})$ and $(\{3\},\{0\})$: they exchange the defenders' roles but both cover $\{0,3\}$ at zero cost. It also contains multiple copies of $(\{0\},\{6\})$, produced by base-specific schedules that become indistinguishable after route information is discarded. Whenever $\mathbf{C}_i=\mathbf{C}_{i'}$ and $\kappa_i=\kappa_{i'}$, the two actions have identical payoff samples for both players and can be merged without changing the payoff set. This exact duplicate removal reduces the simple instance from $169$ actions to its $\binom{10}{1}+\binom{10}{2}=55$ distinct coverage patterns and reduces the general instance from $2704$ actions to $732$ distinct coverage--cost pairs. A further dominance reduction is available when $\mathbf{C}_i=\mathbf{C}_{i'}$ but $\kappa_i<\kappa_{i'}$. The follower's payoff samples are again identical, while Equations~\eqref{eq:risk-aware-guards-signed-payoffs}--\eqref{eq:risk-aware-guards-normalized-payoffs} give $u_L(i,j\mid\xi_k)-u_L(i',j\mid\xi_k)=(\kappa_{i'}-\kappa_i)/(H(\overline r_L-\underline r_L))>0$ for every $j$ and $k$. Transferring probability mass from $i'$ to $i$ therefore leaves the follower's complete optimization problem unchanged and raises every leader sample payoff by the same constant; translation equivariance raises the leader's utility by that constant. Retaining one minimum-cost representative for each coverage vector consequently reduces the general computational model from $732$ to $401$ actions without changing its optimal Stackelberg value. We keep the original normalization constants and a map from each retained representative to its original GUARD action, so every reduced commitment lifts directly to a valid commitment over the original action set.

\paragraph{Optimization Procedure.}
For $\theta_F=0$, we enumerate the polyhedral regions in which each pure follower action is optimal, maximize the leader utility within every feasible region using linear programming for $\theta_L=0$ and smooth concave optimization for $\theta_L>0$, and retain the best regional solution. For $\theta_F>0$, we use the ten-restart implicit-gradient method.

\section{Additional Experimental Results}\label{app:addtional}

\begin{figure*}[t]
\centering
\begin{minipage}{0.49\textwidth}
\centering
\includegraphics[width=\linewidth]{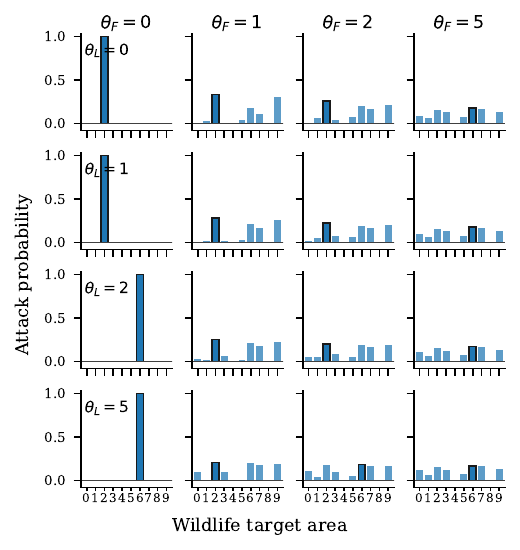}
\textbf{(a) Attacker probabilities: simple schedules}
\end{minipage}\hfill
\begin{minipage}{0.49\textwidth}
\centering
\includegraphics[width=\linewidth]{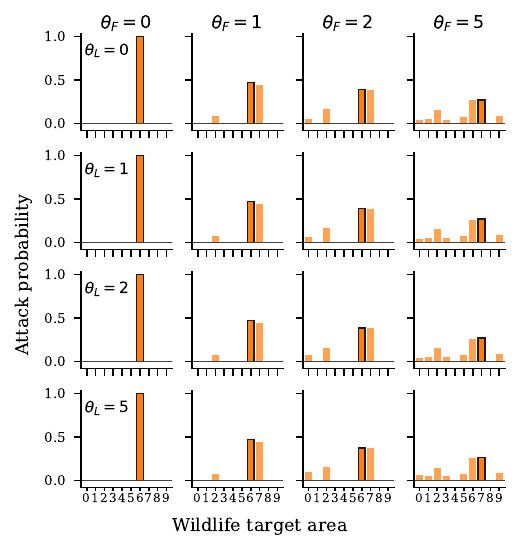}
\textbf{(b) Attacker probabilities: general schedules}
\end{minipage}

\vspace{0.5em}
\begin{minipage}{0.49\textwidth}
\centering
\includegraphics[width=\linewidth]{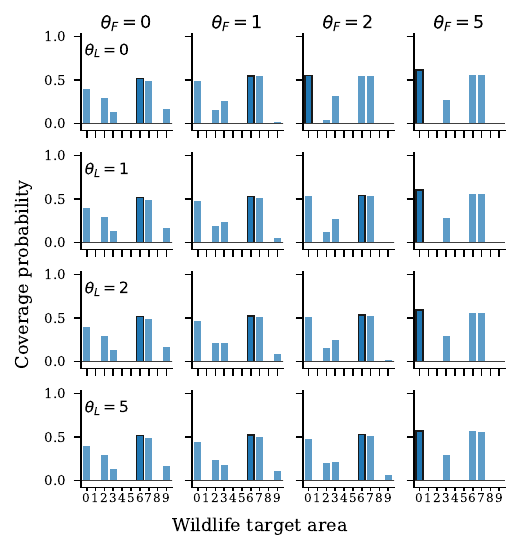}
\textbf{(c) Defender coverage: simple schedules}
\end{minipage}\hfill
\begin{minipage}{0.49\textwidth}
\centering
\includegraphics[width=\linewidth]{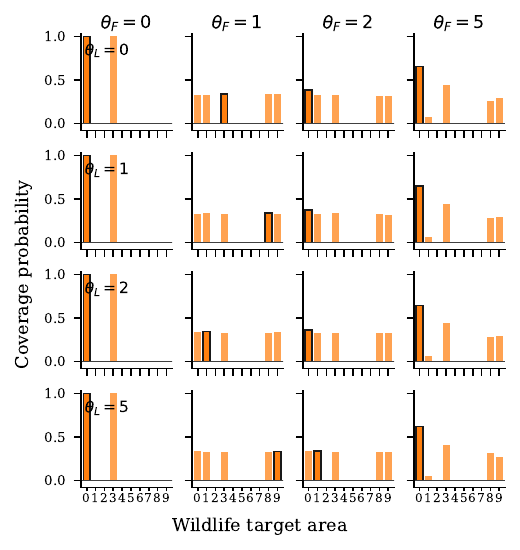}
\textbf{(d) Defender coverage: general schedules}
\end{minipage}
\caption{Attacker target probabilities and defender marginal coverage across the complete risk-parameter grid. Rows vary the defender parameter $\theta_L$, columns vary the attacker parameter $\theta_F$, and bars correspond to the ten wildlife targets.}
\label{fig:risk-aware-guards-appendix-target-profiles}
\end{figure*}

\paragraph{Attacker Response and Defender Coverage.}
Figure~\ref{fig:risk-aware-guards-appendix-target-profiles} reports the complete parameter grid correspodning to Figure~\ref{fig:risk-aware-guards-target-profiles} for \textsc{Risk-Aware-Guards}. In both schedule configurations, the risk-neutral attacker selects a single target, whereas increasing $\theta_F$ distributes attack probability across more targets. Averaged over $\theta_L$, the largest attack probability decreases from $1.000$ at $\theta_F=0$ to $0.268$, $0.218$, and $0.175$ under simple schedules, and to $0.472$, $0.386$, and $0.268$ under general schedules, for $\theta_F=1,2,5$, respectively. Under simple schedules, the defender correspondingly concentrates coverage on the high-frequency wildlife targets, particularly targets $0$, $3$, $6$, and $7$. General schedules permit broader simultaneous coverage; after the initial redistribution at $\theta_F=1$, increasing $\theta_F$ similarly shifts coverage toward these targets. Across both schedules, $\theta_L$ primarily changes the magnitude of the coverage reallocation, while the qualitative effect of $\theta_F$ remains consistent.

\end{document}